\documentclass{amsproc}
\usepackage{epstopdf}
\usepackage{graphicx}
\usepackage{amssymb}
\usepackage{empheq}
\usepackage{cases}
\usepackage{amsthm,amsmath}
\usepackage{caption,lipsum}
\newcommand{\eps}{\varepsilon}
\newcommand{\bN}{\mathbb{N}}
\newcommand{\bR}{\mathbb{R}}

\newcommand{\fe}{\mathrm{e}}

\newcommand{\bx}{\mathbf{x}}
\newcommand{\ud}{\mathrm{d}}
\newcommand{\sech}{\mathrm{sech}}
\newtheorem{prop}{Proposition}[section]
\newtheorem{theorem}{Theorem}[section]
\newtheorem{remark}{Remark}[section]

\newcommand{\be}{\begin{equation}}
\newcommand{\ee}{\end{equation}}
\newcommand{\ba}{\begin{array}}
\newcommand{\ea}{\end{array}}
\newcommand{\bea}{\begin{eqnarray}}
\newcommand{\eea}{\end{eqnarray}}
\newcommand{\beas}{\begin{eqnarray*}}
\newcommand{\eeas}{\end{eqnarray*}}
\input psfig.sty

\numberwithin{equation}{section}

\begin{document}

\title[Algorithms for multichannel solutions in NLS]{On multichannel solutions of nonlinear Schr\"{o}dinger equations: algorithm, analysis and numerical explorations}

\author{Avy Soffer}
\address{Department of Mathematics, Rutgers University, New Jersey 08854}
\email{soffer@math.rutgers.edu}

\author{Xiaofei Zhao}
\address{Department of Mathematics, National University of Singapore, Singapore 119076}
\email{zhxfnus@gmail.com}

\subjclass[2000]{Primary}

\begin{abstract}
We apply the method of modulation equations to numerically solve the NLS with multichannel dynamics, given by a trapped localized state and radiation.
This approach employs the modulation theory of Soffer-Weinstein, which gives a system of ODE's coupled to the radiation term, which is valid for all times.
We comment on the differences of this method from the well-known method of collective coordinates.
\end{abstract}

\maketitle

\section{Introduction}
\label{sec: intr}
A wide class of conservative nonlinear dispersive equations, such as nonlinear Schr\"{o}dinger equations and Korteweg-de Vries equations (See e.g. the reviews \cite{CSC,Sof} ), admit solutions with more than one channel or the multichannel solutions, which means that the
asymptotic behavior of the solution is given by a linear combination of a localized (in space), periodic (in time) wave (solitary or standing wave) and a dispersive part.
The multichannel solutions are important and useful in both theoretical analysis and applications. For example, they occur frequently in the nonlinear scattering theory in the study of nonintegrable equations and the design of absorbing boundary conditions for the partial differential equations. They also frequently appear in various nonlinear systems, such as quantum waves and nonlinear optics. In general, the dynamics is complicated due to the interaction between the coherent structures and the radiation around.

Among the analytical approaches to study the multichannel solutions,
the collective coordinate approach is one of the most useful tools, see a detailed review in \cite{Dawes}. The collective coordinate approach, see e.g. \cite{Alawi,BBB2,Alamoudi,Collective2,Nishida,Quintero,374, Collective},
approximates the solution at any given time, by the nearest soliton structure possible; this gives ODE's for the dynamics of the soliton(s) parameters, which decouple from the rest of the system. This method is very effective in predicting the dynamics of solitons in nonlinear systems \cite{Morales4,Morales3,Morales2,Morales1} and has some other applications like in \cite{Zamora}, but it does not take the radiations into consideration. In 1990, A. Soffer and M. I. Weinstein rigorously established the mathematical theory and derived the governing dynamical equations for the multichannel solutions in general system when they were studying the scattering in nonintegrable equations \cite{Soffer1}. The governing equations of the multichannel solutions can fully describe the dynamics of the solitary wave and the dispersive part simultaneously, and they are known as modulation equations in literature.
However, little numerical aspects have been done to the study of the multichannel solutions by modulation equations approach so far. In this work, we focus on the multichannel solutions in the nonlinear Schr\"{o}dinger (NLS) equations in $d$ dimensions($d=1,2,3$), i.e.
\begin{eqnarray}
&i\partial_t\Phi(\mathbf{x},t)=\left[-\Delta+V(\mathbf{x})+\lambda|\Phi|^{2m}\right]\Phi(\mathbf{x},t),\quad \mathbf{x}\in\bR^d,\ t>0,\label{nls}\\
&\Phi(\mathbf{x},0)=\Phi_0(\mathbf{x}),\quad \mathbf{x}\in\bR^d,
\end{eqnarray}
where $m\in\bN$, $\lambda\in\bR$ and $0<|\lambda|\ll1$, $\Phi_0(\mathbf{x})$ is the given initial data and $V(\mathbf{x})$ is the real-valued potential function such that $-\Delta+V$ has continuous spectrum. It is well-known that mass of system $M$ and the Hamiltonian (energy) of the system $H$ are conserved, i.e.
\begin{eqnarray}
M(t)&:=&\int_{\bR^d}\left|\Phi(\mathbf{x},t)\right|^2\ud \mathbf{x}\equiv M(0),\\
H(t)&:=&\int_{\bR^d}\left[\left|\nabla\Phi(\mathbf{x},t)\right|^2+V(\mathbf{x})\left|\Phi(\mathbf{x},t)\right|^2+\frac{\lambda}{m+1}\left|\Phi(\mathbf{x},t)\right|^{2m+2}
\right]\ud \mathbf{x}\nonumber\\
&\equiv& H(0).\label{E&M}
\end{eqnarray}

Based on the modulation equations of the NLS provided in \cite{Soffer1}, accurate numerical algorithms are proposed, analyzed and applied to explore the multichannel solutions numerically.
In the study, the key and necessary step for the numerical algorithm of modulation equations is to solve the nonlinear eigenvalue problems of the NLS in the case of continuous spectrum, which to our best knowledge, no applicable numerical methods are available in literature. For this problem, the existing numerical methods only consider either the ground state of the NLS or the excited states in case of discrete spectrum \cite{Bao1,BaoDu,Feit}. Thus, before we design the algorithm for the modulation equations, we propose and analyze a new effective numerical method for solving the eigenvalue problem of the NLS with continuous spectrum.

The rest of the paper is organized as follows. In Section \ref{sec: review}, we shall give a brief review of the multichannel solutions of the NLS. In Section \ref{sec: eigen pro}, we shall propose and analyze the algorithm for the nonlinear eigenvalue problem of the NLS. The algorithm for the modulation equations is given in Section \ref{sec: modulate eqns} followed by numerical explorations.

\section{Review of multichannel solutions}\label{sec: review}
In this section, we shall formally derive the modulation equations of the NLS and review some related mathematical theories for the readers' convenience. Comments on the differences of this method from the well-known method of collective coordinates are given in the end.

\subsection{Formal derivations of modulation equations}
Based on the simple physical observation that
if one starts with the linear Schr\"{o}dinger equation which describes a bound
state and a dispersive wave (corresponding to the continuous spectral part of
the Hamiltonian), then the qualitative behavior should not change that much in
response to a small nonlinear and Hamiltonian perturbation in the dynamics, i.e.
we should still see a localized part which decouples after a long time from the
dispersive part, we take the ansatz of the solution of (\ref{nls}) \cite{Soffer1} as
\begin{eqnarray}
&\Phi(\mathbf{x},t)=\fe^{-i\theta(t)}\left(\psi_{E(t)}(\mathbf{x})+\phi(\mathbf{x},t)\right),\quad\mathbf{x}\in\mathbf{R}^d,\ \,t\geq0,\label{Ansatz}\\
&\theta(t)=\int_0^tE(s)\ud s-\gamma(t),\quad t\geq0,
\end{eqnarray}
with initial conditions
\begin{eqnarray}
&&\Phi_0(\bx)=\fe^{-i\theta_0}\left(\psi_{E_0}(\bx)+\phi_0(\bx)\right),\nonumber\\
&&E(0)=E_0,\quad \gamma(0)=\gamma_0,\quad \phi(\bx,t=0)=\phi_0(\bx).\label{modulation: ini}
\end{eqnarray}
Here, $\psi_E(\bx)$ is the nonlinear bound state of NLS (\ref{nls}) with eigenvalue $E$, i.e.
\begin{eqnarray}
&\left[-\Delta+V(\mathbf{x})+\lambda\psi_E^{2m}(\mathbf{x})\right]\psi_E(\mathbf{x})=E\psi_E(\mathbf{x}),\quad
\mathbf{x}\in\bR^d,\label{eigenfunction}\\
& \psi_E\in H^2(\bR^d),\qquad \psi_E>0,\nonumber
\end{eqnarray}
which is the localized part in the multichannel solution.
$\phi$ is the dispersive wave and $\gamma\in\bR$ is the frequency.
Also, we assume the orthogonality condition:
\begin{equation}\label{ansatz: orth}
\left<\psi_{E_0},\phi_0\right>=0,\quad \frac{\ud}{\ud t}\left<\psi_{E_0},\phi(t)\right>=0,
\end{equation}
where the inner product $<u,v>:=\int_{\bR^d}\bar{u}v\,\ud \mathbf{x}$ for functions $u$ and $v$, with $\bar{u}$ the complex conjugate of $u$. Condition (\ref{ansatz: orth}) immediately implies that $\left<\psi_{E_0},\phi(t)\right>=0$.

Plugging ansatz (\ref{Ansatz}) into the NLS (\ref{nls}) and noting (\ref{eigenfunction}), we get
\begin{eqnarray}
i\partial_t\phi(\mathbf{x},t)=&\left[-\Delta+V-E(t)+\dot{\gamma}(t)\right]\phi
+\lambda\left|\psi_{E(t)}+\phi\right|^{2m}\left(\psi_{E(t)}+\phi\right)\label{disper eqn}\\
&-\lambda\psi_{E(t)}^{2m+1}+\dot{\gamma}(t)\psi_{E(t)}-i\dot{E}(t)\partial_E\psi_{E(t)},\nonumber
\end{eqnarray}
where from (\ref{eigenfunction}) we have
\begin{equation}\label{dE psi}
\left[-\Delta+V+(2m+1)\lambda\psi_E^{2m}-E\right]\partial_E\psi_E=\psi_E.
\end{equation}
Denote
\begin{subequations}
\begin{eqnarray}
&F_1=\dot{\gamma}\psi_E-i\dot{E}\partial_E\psi_E,\label{F1 def}\\
&F_2=\lambda\left|\psi_E+\phi\right|^{2m}\left(\psi_E+\phi\right)-\lambda\psi_E^{2m+1}-\lambda\psi_{E_0}^{2m}\phi,
\end{eqnarray}
\end{subequations}
then (\ref{disper eqn}) becomes,
\begin{eqnarray*}
i\partial_t\phi(\mathbf{x},t)=&\left(-\Delta+V+\lambda\psi_{E_0}^{2m}\right)\phi
+\left(\dot{\gamma}-E\right)\phi+F_1+F_2.
\end{eqnarray*}
Taking the inner product on both sides of the above equation with $\psi_{E_0}$, we get
\begin{eqnarray*}
i\left<\partial_t\phi,\psi_{E_0}\right>=&\left<(-\Delta+V+\lambda\psi_{E_0}^{2m})\phi,\psi_{E_0}\right>+
\left(\dot{\gamma}-E\right)\left<\phi,\psi_{E_0}\right>\\
&+\left<F_1+F_2,\psi_{E_0}\right>.
\end{eqnarray*}
Using (\ref{eigenfunction}) and the orthogonal condition (\ref{ansatz: orth}), then noting (\ref{F1 def}) we get
$$\left<F_2,\psi_{E_0}\right>=-\left<F_1,\psi_{E_0}\right>=-\dot{\gamma}\left<\psi_E,\psi_{E_0}\right>-i\dot{E}\left<\partial_E\psi_E,\psi_{E_0}\right>.$$
Taking the real part and imaginary part of the above equation, respectively, we get
\begin{eqnarray*}
\dot{\gamma}(t)=-\frac{\mathfrak{R}\left<F_2,\psi_{E_0}\right>}{\left<\psi_E,\psi_{E_0}\right>},\qquad
\dot{E}(t)=-\frac{\mathfrak{I}\left<F_2,\psi_{E_0}\right>}{\left<\partial_E\psi_E,\psi_{E_0}\right>},
\end{eqnarray*}
where $\mathfrak{R}z$ and $\mathfrak{I}z$ denote the real part and imaginary part of a complex number $z$, respectively. Together with (\ref{disper eqn}), we get the \emph{modulation equations} for the dynamics of multichannel solutions as the coupled system
\begin{subequations}\label{modulation}
\begin{eqnarray}
&i\partial_t\phi(\mathbf{x},t)=\left[-\Delta+V-E(t)+\dot{\gamma}(t)\right]\phi
+\lambda\left|\psi_{E(t)}+\phi\right|^{2m}\left(\psi_{E(t)}+\phi\right)\label{mudulated: disper}\\
&\qquad\qquad\quad\,-\lambda\psi_{E(t)}^{2m+1}+\dot{\gamma}(t)\psi_{E(t)}-i\dot{E}(t)\partial_E\psi_{E(t)},\quad \mathbf{x}\in\bR^d,\ \, t>0,\nonumber\\
&\dot{\gamma}(t)=-\frac{\mathfrak{R}\left<F_2,\psi_{E_0}\right>}{\left<\psi_E,\psi_{E_0}\right>},\quad t>0,\\
&\dot{E}(t)=-\frac{\mathfrak{I}\left<F_2,\psi_{E_0}\right>}{\left<\partial_E\psi_E,\psi_{E_0}\right>},\quad t>0,
\end{eqnarray}
\end{subequations}
with initial conditions (\ref{modulation: ini}). Components $E$ and $\gamma$ are sometimes referred as \emph{collective coordinates} in physics, and the dispersive term $\phi$ are related to \emph{radiations}.

\subsection{Review on theoretical results}
Here always assume the linear operator $- \Delta+V$ has continuous spectrum but with exactly one bound state (isolated eigenvalue) on $L^2(\mathbb{R}^d)$  with strictly negative eigenvalue $E_*$. In fact, the nonlinear eigenvalue problem (\ref{eigenfunction}) and the coupled modulation equations (\ref{modulation}) have been well-studied by A. Soffer and M. I. Weinstein in \cite{Soffer1}. Here we will briefly state some main results established there on the eigenvalue problem and the modulation equations. For the detailed theoretical statements and proofs, we refer the readers to \cite{Soffer1}.

\emph{For the nonlinear eigenvalue problem}

For any $E\in(E_*,0)$ if $\lambda>0$ (defocusing case), and for any $E<E_*$ if $\lambda<0$ (focusing case),
the nonlinear eigenvalue problem (\ref{eigenfunction}) has a unique positive solutions in $H^2(\mathbb{R}^d)$.

\emph{For the modulation equations}

(Global well-posedness) For initial conditions $\phi(\bx,0)=\phi_0(\bx)\in H^1(\mathbb{R}^d)$, $\gamma(0)=\gamma_0\in\mathbb{R}$, and $E(0)=E_0\in\mathbb{R}$, where $E_0\in(E_*,0)$ if $\lambda>0$ and $E_0<E_*$ if $\lambda<0$, the modulation equations (\ref{modulation}) has a unique solution $E(t),V(t)\in C^1[0,\infty)$ and $\phi\in C([0, \infty); H^1).$

(Dynamical property) For $0<t<\infty$, the solutions $E(t)\in (E_*,0)$ if $\lambda>0$ and $E(t)<E_*$ if $\lambda<0$. As $t\to \infty,$ there exist two constants $\tilde{E}, \tilde{\gamma}$, such that $E(t)\to \tilde{E}$, $\gamma(t)\to\tilde{\gamma}$.

\subsection{ Relation of modulation equations to collective coordinates}
The method of modulation equations described above is closely related to the method of collective coordinates.
In fact, it was proven in \cite{Soffer1} that if the solution at any time is close to a soliton, then it can be written as a soliton plus small remainder which is also orthogonal in the above sense. Hence, both methods give a decomposition with small corrections.
However, the method of the modulation equations also give a PDE for the radiation term, and coupling between the radiation and the ODE's.
Hence the modulation equations approach allows : (i) Control of the error in the ODE's.
(ii) Allows approximating the effect of radiation on the soliton dynamics, either exactly, or by using a good approximation of the coupling term.
(iii) When the radiation effect is critical, as in dissipation mediated processes by the radiation \cite{Soffer2,Soffer3}, one can derive the leading dissipation (radiation mediated!) term from the coupled equations, and find the leading behavior for processes in which a soliton changes state, for example from excited to ground state. (see \cite{Soffer3, Soffer4}).
(iv) Resolving the soliton part as it arrives at the boundary of the domain of computation. This can not be handled by absorbing boundaries \cite{SStu}.
(v) Allows the rigorous asymptotic stability and scattering over arbitrary large time intervals.

\section{On the nonlinear eigenvalue problem}\label{sec: eigen pro}
In order to solve the modulation equations (\ref{modulation}) at time $t$, we notice that we need to compute the bound state $\psi_{E(t)}$ which is the solution of the nonlinear eigenvalue problem (\ref{eigenfunction}). Thus, in this section, we shall propose and analyze an efficient algorithm to compute the bound state of (\ref{eigenfunction}) who has continuous spectrum.

\subsection{Numerical method}
For a given $E$ within the spectrum, since the bound state decays very fast to zero at far field, we truncate problem (\ref{eigenfunction}) onto a finite interval $\Omega=[-L,L]^d$ and impose the homogeneous Dirichlet boundary condition for numerical aspects, i.e.
\begin{subequations}\label{eigen trun}
\begin{eqnarray}
&\left[-\Delta+V(\mathbf{x})+\lambda\psi_E^{2m}(\mathbf{x})\right]\psi_E(\mathbf{x})=E\psi_E(\mathbf{x}),\quad
\mathbf{x}\in\Omega,\label{eigen trun a}\\
& \psi_E(\mathbf{x})=0,\quad \mathbf{x}\in\partial\Omega.\label{eigen trun b}
\end{eqnarray}
\end{subequations}

The algorithm reads as the following. Denote $\psi^{(n)}_E(n=0,1,\ldots,)$ as the approximation to $\psi_E$ and suppose $\psi^{(0)}_E$ is the initial guess, then
$\psi_E^{(n+1)}$ is updated as:

\emph{Step 1} Find the ground state of the Hamiltonian functional
\begin{equation*}
H^n(\psi)=\int_\Omega\left(|\nabla\psi|^2+V|\psi|^2+\lambda\left|\psi_E^{(n)}\right|^{2m}\cdot|\psi|^2\right)\ud \mathbf{x},
\end{equation*}
in the unit sphere of $L^2(\Omega)$. Denote the solution as
\begin{equation}\label{algor: eigen1}
\widetilde{\psi}_E^{(n+1)}:= \arg \min\{H^n(\psi):\psi\in L^2(\Omega),\ \|\psi\|_{L^2}=1\}.
\end{equation}

\emph{Step 2} Scale the ground state $\widetilde{\psi}_E^{(n+1)}$ according to the energy $E$. That is to find the scaling constant $c^n$ such that
\begin{equation}\label{algor: eigen2}
\psi_E^{(n+1)}:=c^n\widetilde{\psi}_E^{(n+1)},
\end{equation}
satisfies
\begin{eqnarray*}
\int_\Omega\left[\left|\nabla\psi_E^{(n+1)}\right|^2+V\left|\psi_E^{(n+1)}\right|^2+\lambda\left|\psi_E^{(n+1)}\right|^{2m+2}\right]\ud \mathbf{x}=E\left\|\psi_E^{(n+1)}\right\|_{L^2}^2,
\end{eqnarray*}
where we obtain
\begin{equation*}
c^n=\left|\frac{E-\int_\Omega\left[\left|\nabla\widetilde{\psi}_E^{(n+1)}\right|^2
+V\left|\widetilde{\psi}_E^{(n+1)}\right|^2\right]\ud \mathbf{x}}
{\lambda\int_\Omega\left|\widetilde{\psi}_E^{(n+1)}\right|^{2m+2}\ud\mathbf{x}}\right|^{\frac{1}{2m}}.
\end{equation*}

For the first step, we use the normalized gradient flow method with a backward Euler sine pseudospectral discretization \cite{Bao1,BaoDu} to get the ground state $\widetilde{\psi}_E^{(n+1)}$. For the second step, we use the standard sine pseudospectral discretization \cite{book-sp-GO, Shen} for the spatial derivative and integrations. Once $\psi_E$ is obtained, $\partial_E\psi_E$ can be found out from (\ref{dE psi}) with zero boundary conditions on $\Omega$ by the sine pseudospectral discretization.

\subsection{Convergence analysis}
For the proposed algorithm (\ref{algor: eigen1})-(\ref{algor: eigen2}), we have the convergence result stated as the following theorem.
\begin{theorem}\label{thm}
For the iteration algorithm (\ref{algor: eigen1})-(\ref{algor: eigen2}) for solving the eigenvalue problem (\ref{eigen trun}), if $\psi_E^{(0)}$ is sufficiently close to $\psi_E$ in $H^1(\Omega)$, we have
\begin{equation}
\left\|\psi_E^{(n)}-\psi_E\right\|_{H^1(\Omega)}\rightarrow0, \qquad n\rightarrow\infty.
\end{equation}
\end{theorem}
The proof of the convergence theorem is given by the following steps.
Let $\psi_0$ with $\|\psi_0\|_{L^2}=1$ be the ground state of $-\Delta+V(\mathbf{x})$ on $L^2(\mathbf{R})$ with the eigenvalue $E_*<0$. For simplicity, we work in the one space dimension case, i.e. $d=1,\ \bx=x$. Furthermore, we assume that $E$ is in some small neighborhood of $E_*$ to be determined, and we use the $\psi_E^0=\psi_0$ as the initial guess for the iteration scheme.
For general cases, the argument can proceed in a similar way.
Define the functional, which is the scaling step used in (\ref{algor: eigen2})
$$\eta_E: \psi(\in H^1)\to \mathbb{R}^1,$$
$$\eta_E(\psi):=\frac{E-\int|\psi|^2\ud x-\int\left(|\nabla \psi|^2+V|\psi|^2\right)\ud x}{\lambda\int|\psi|^{2m+2}\ud x},
\quad \mbox{and}\quad \widetilde{\eta}_E(\psi):=\left|\eta_E(\psi)\right|^{\frac{1}{2m}}.$$

 \begin{prop}\label{prop}
 The following three facts are true.

  a) $\eta_E(\psi_0)=\frac{E-E_*}{\lambda\int |\psi_0|^{2m+2}\ud x}.$

  b) Suppose $\|\psi-\psi_0\|_{H^1}\leq b\ll1,$ then
 $$|\eta_E(\psi)|\leq |\eta_E(\psi_0)|+C|E-E_*|\|\psi-\psi_0\|_{H^1}+C\|\psi-\psi_0\|_{H^1}\|\psi-\psi_0\|_{L^2}.$$

 c) For $\|\psi-\psi_0\|_{H^1}+\|\psi'-\psi_0\|_{H^1}\leq 2b$,
 $$|\eta_E(\psi)-\eta_E(\eta')|\leq C|E-E_*|\|\psi-\psi'\|_{H^1}+O(\|\psi-\psi_0\|_{H^1}+\|\psi'-\psi_0\|_{H^1})\|\psi-\psi'\|_{L^2}.$$

\end{prop}

\emph{Proof of Proposition \ref{prop}}:

a) Since $\psi_0$ is normalized to $1$ in $L^2$ and $-\Delta \psi_0+V\psi_0=E_*\psi_0,$ so
\begin{equation*}
\eta_E(\psi_0)=\frac{ E-\int\left(|\nabla\psi_0|^2+V|\psi_0|^2\right)\ud x}{\lambda\int |\psi_0|^{2m+2}\ud x}=\frac{E-E_*}
{\lambda\int\psi_0^{2m+2}\ud x}.
\end{equation*}
b) Expand $\eta_E(\psi)$ around $\psi_0$:
\begin{eqnarray*}
\eta_E(\psi)=&\eta_E(\psi_0)+\frac{\delta\eta}{\delta\psi}\big|_{\psi=\psi_0}(\psi-\psi_0)
+\frac{\delta^2\eta}{\delta\psi_i\delta\psi_j}\big|_{\psi=\psi_0}(\psi_i-\psi_0)(\psi_j-\psi_0)\\
&+O((\psi-\psi_0)^2),
\end{eqnarray*}
where $\psi_i=\psi$ or $\psi^*$. At the `point' $\psi_0$, by definition of the ground state, we have
\begin{eqnarray*}
\frac{\delta\eta}{\delta\psi}\big|_{\psi=\psi_0}&&=\frac{1}{\lambda\int|\psi_0|^{2m+2}\ud x}\frac{\delta}{\delta\psi}
\{E(\psi,\psi)-(\psi,H_0\psi)\}\\
&&\quad+\frac{E(\psi_0,\psi_0)-(\psi_0,H_0\psi_0)}{\lambda(\int|\psi_0|^{2m+2}\ud x)^2}
\frac{\delta}{\delta\psi}\int|\psi|^{2m+2}\ud x\\
&&=\frac{(E-E_*)<\psi_0|+\frac{E-E_*}{\int|\psi_0|^{2m+2}\ud x}C_m<|\psi_0|^m\psi_0^*|}{\lambda\int|\psi_0|^{2m+2}\ud x},
\end{eqnarray*}
where $H_0=-\Delta+V+\lambda\psi_0^{2m}$, and $<\phi|$ stands for the operator $<\phi|f=\int\phi^* f\ud x.$ Therefore,
$$\left|\frac{\delta\eta}{\delta\psi}\big|_{\psi=\psi_0}(\psi-\psi_0)\right|\leq \frac{C_m+1}{|\lambda|\int|\psi_0|^{2m+2}\ud x}
|E-E_*|\left\|\psi\right\|_{L^2}\|\psi-\psi_0\|_{H^1}.$$
The higher order terms are similarly controlled, where we use that
$$\|\psi_i-\psi_0\|_{L^\infty}\leq C\|\psi_i-\psi_0\|_{H^1}.$$

c) Using the expansion in b), we have that
$$\eta_E(\psi)-\eta_E(\psi')=\frac{\delta\eta}{\delta\psi}\big|_{\psi=\psi_0}(\delta\psi-\delta\psi')+O(\delta\psi^2-\delta(\psi')^2),$$
where $\delta\psi=\psi-\psi_0$, $\delta\psi'=\psi'-\psi_0.$ Hence,
\begin{eqnarray*}
&\quad\left|\eta_E(\psi)-\eta_E(\psi')\right|\\
&\leq \frac{C}{|\lambda|\int |\psi_0|^{2m+2}\ud x}\Big[|E-E_*|(\|\psi\|_{H^1}+\|\psi'\|_{H^1}+1)\|\psi-\psi'\|_{L^2}\\
&\quad+
\|\psi-\psi'\|_{L^2}O(\|\psi-\psi_0\|_{H^1}+\|\psi'-\psi_0\|_{H^1})\Big].
\end{eqnarray*}
Then the proof of the proposition is completed. \begin{flushright} $\square$\end{flushright}

Now, we can proceed to the proof of the convergence of the scheme, which is similar to the way used in \cite{proof}.

\emph{Proof of Theorem \ref{thm}}
We need to show that the mapping defined by the iteration scheme is a strict contracting mapping for $|E-E_*|$ sufficiently small, on a complete metric space containing $\psi_0$ as an internal point. So we choose the metric space $\mathcal{M}_{0,\varepsilon}$ to be
$$\mathcal{M}_{0,\varepsilon}:=\left\{\psi\in H^1|\,\left\|\psi-\psi_0\right\|\leq \varepsilon,\ \|\psi\|_{L^2}=1 \right\}.$$
Now, $\psi_0\in \mathcal{M}_{0,\varepsilon}$. Hence, for any $\psi\in \mathcal{M}_{0,\varepsilon}$, we have that
\begin{eqnarray*}
|\lambda\eta_E(\psi)|&&\leq|\lambda||\eta_E(\psi)-\eta_E(\psi_0)|+\lambda\eta_E(\psi_0)\\
&&\leq C(\psi_0,\varepsilon)|E-E_*|\|\psi-\psi_0\|_{L^2}+C(\psi_0)|E-E_*|\\
&&\quad+C\|\psi-\psi_0\|_{H^1}\|\psi-\psi_0\|_{L^2}\\
&&\leq C(\psi_0,\varepsilon)|E-E_*|\varepsilon+C(\psi_0)|E-E_*|+O(\varepsilon^2).
\end{eqnarray*}
Define the map $S: \mathcal{M}_{0,\varepsilon}\to H^1$ by
$$S(\psi):=\frac{P_g^{\psi}\psi_0}{\|P_g^{\psi}\psi_0\|_{L^2}},$$
where $P_g^{\psi}$ denotes the projection on the ground state of the Hamiltonian:
$$H_\psi:=-\Delta+V+\lambda\eta_E(\psi)|\psi|^{2m}.$$
Clearly, $\|S(\psi)\|_{L^2}=1$, and $S(\psi)\in H^1$, since the ground state of $-\Delta+V+\lambda f(x)$ is smooth and exponentially decaying by general theory for such $f(x)$. Our $f(x)$ is bounded in $H^1$, and therefore in every $L^p$. It is also small since $\lambda\eta_E(\psi)$ is small for small $\varepsilon$. So it remains to estimate the $H^1$ norm of such $S(\psi)$. Let's first compute the $H^1$ norm of $S(\psi_0)$:
\begin{eqnarray*}
&\quad\left\|\|S(\psi_0)-\frac{1}{2\pi i}\oint_{\Gamma}\frac{\ud z}{H_0-z}\psi_0\right\|_{H^1}=\left\|\frac{1}{2\pi i}
\oint_{\Gamma}\ud z(H_{\psi_0}-z)^{-1}\psi_0\right\|\\
&=\left\|\frac{1}{2\pi}\oint_{\Gamma}\ud z (-\Delta+V+\lambda\eta_E(\psi_0)|\psi_0|^{2m}-z)^{-1}\psi_0\right\|_{H^1}\\
&=\frac{1}{2\pi}\left\|\oint_\Gamma\ud z(-\Delta+V-z)^{-1}\frac{|E-E_*|}{\int|\psi_0|^{2m+2}\ud x}|\psi_0|^{2m}H_{\psi_0}\psi_0\right\|_{H^1},
\end{eqnarray*}
where $\Gamma$ is a small circle of radius $r$ around $E_*$ and $E^1$ the eigenvalue of $H_{\psi_0}$.
By choosing $E$ sufficiently close to $E_*$, we can insure that for small circle of radius $r$, the eigenvalue $E^1$ is close to $E_*$, i.e. $|E^1-E_*|<\frac{r}{2}$, and $E_*+r<0$. A direct estimate then gives:
$$\|S(\psi_0)-\psi_0\|_{H^1}\leq\frac{1}{2\pi}2\pi r C_{\psi_0}|E-E_*|<\varepsilon,$$
by choosing $|E-E_*|$ sufficiently small, where
$$C_{\psi_0}=\sup_{z\in\Gamma}\left(\left\|\frac{1}{H_0-z}\right\|+\left\|\frac{H_0}{H_0-z}\right\|\right)
\left\|\frac{1}{H_0+\lambda\eta(\psi_0)|\psi_0|^{2m}-z}\right\|
\frac{1}{\int|\psi_0|^{2m+2}\ud x}.
$$
Hence $S(\psi_0)\subset\mathcal{M}_{0,\varepsilon}$ for all sufficiently small $|E-E_*|$. A similar computation gives
$$\left\|S(\psi)-S(\psi')\right\|_{H^1}\leq C|E-E_*|\|\psi-\psi'\|_{H^1}<(1-\delta)\|\psi-\psi'\|_{H^1},$$
for all $\psi,\psi'\in \mathcal{M}_{0,\varepsilon}$. This proves the contraction of the mapping and so does the convergence of the scheme.\begin{flushright} $\square$\end{flushright}

\subsection{Numerical results}\label{subsec: eigen numres}
Here, we present the numerical results of using (\ref{algor: eigen1})-(\ref{algor: eigen2}) to solve (\ref{eigen trun}). For simplicity, we choose an one-dimensional example, i.e. $d=1,\ \mathbf{x}=x$. Take the potential as
$$V(x)=-2b^2\sech^2(bx),$$
with parameter $b$ chosen not too large such that the linear operator $-\Delta+V$ has a unique negative eigenvalue $E_*=-b^2$ with the corresponding eigenstate
$$\psi_0(x)=\sech(bx).$$

\begin{figure}[t!]
$$\begin{array}{cc}
\psfig{figure=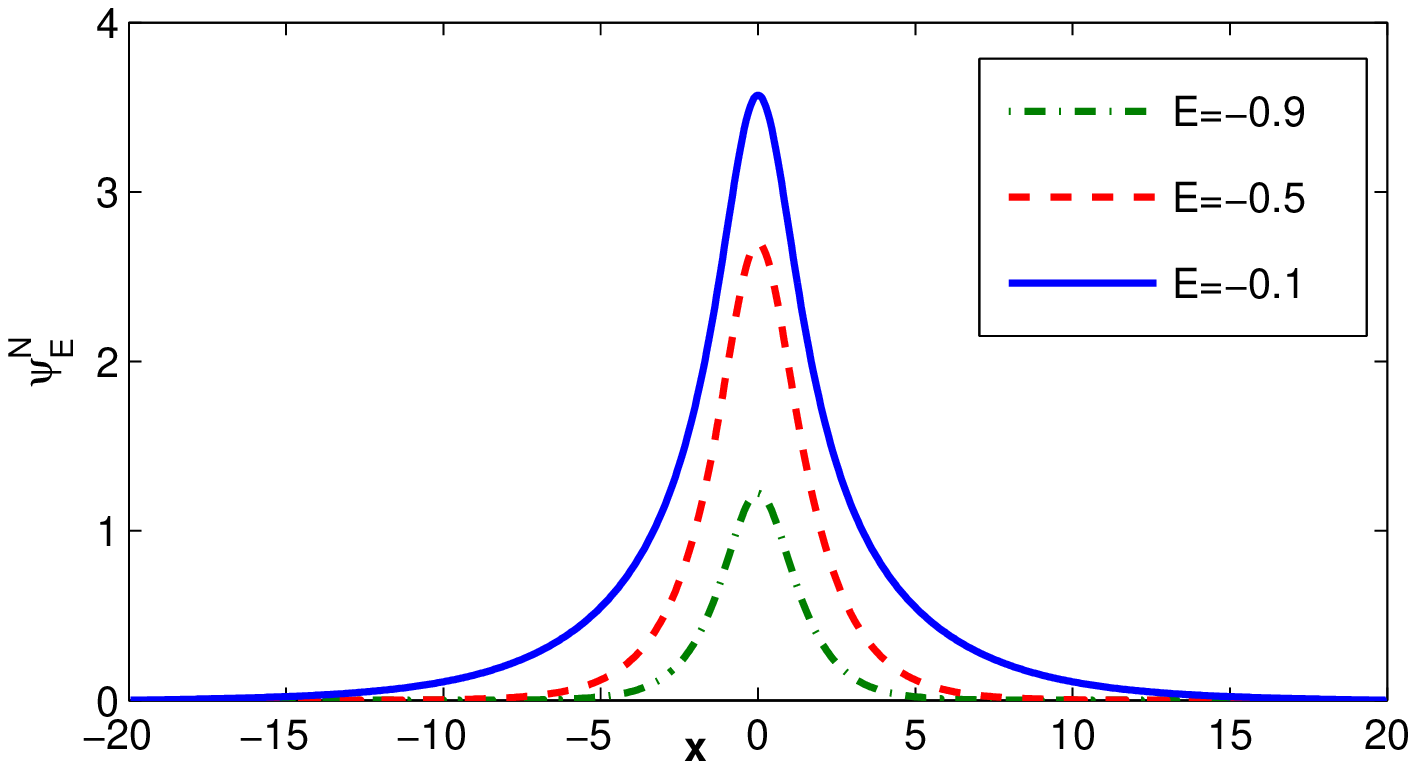,height=5cm,width=6.3cm}&
\psfig{figure=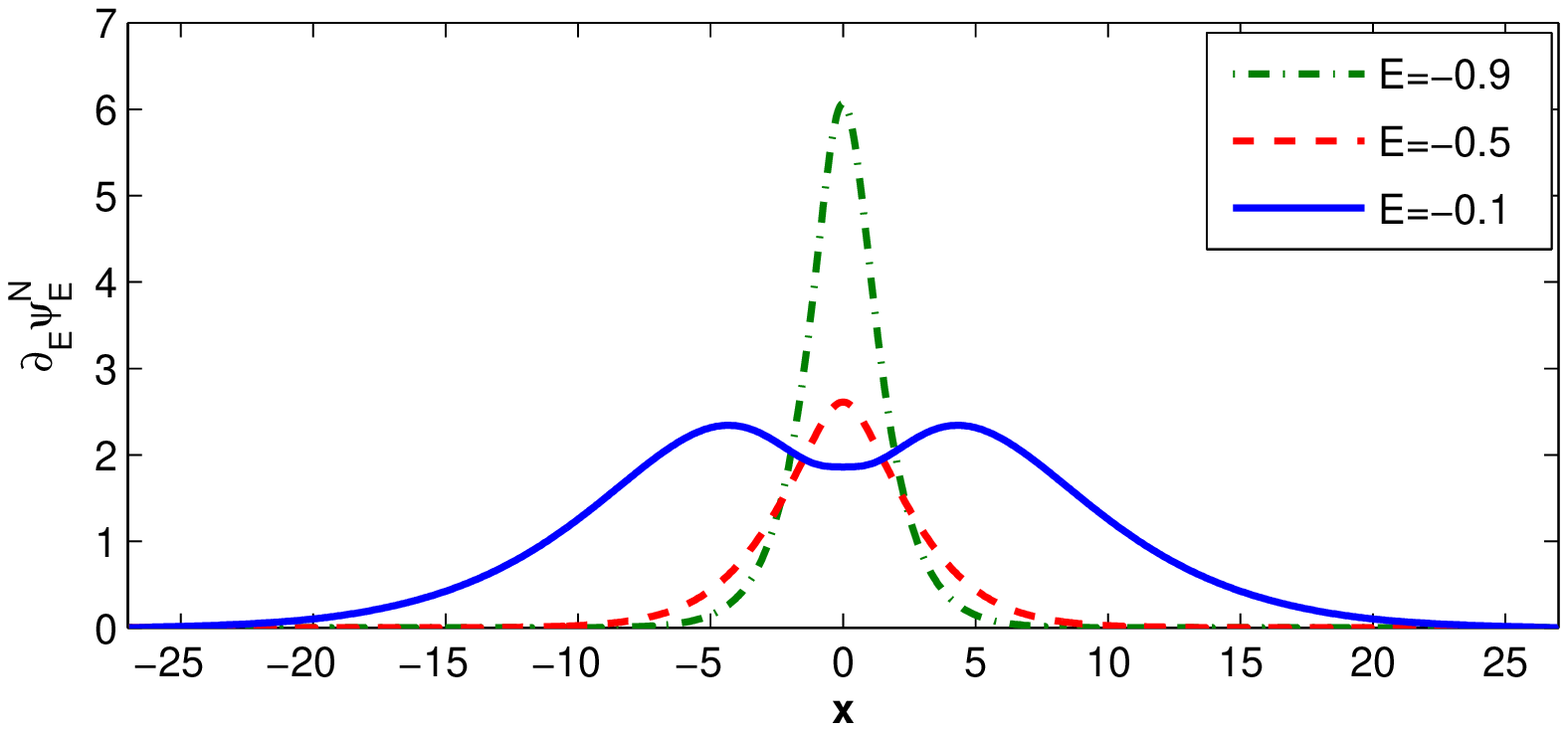,height=5cm,width=6.3cm}
\end{array}$$
\caption{Profiles of the bound state $\psi_E^{(N)}$ and $\partial_E\psi_E^{(N)}$ at different E. }\label{fig}
\end{figure}

Choose $b=1$, $\lambda=0.1$ and $m=1$ (the  Gross-Pitaevskii equation \cite{Bao1}) in the power nonlinearity, and choose the domain $\Omega=[-20,20]$, i.e. $L=20$, which is large enough to neglect the boundary truncation error. Taking the spatial mesh size $h=1/16$, for an arbitrary $E\in(E_*,0)$, we compute the bound state of (\ref{eigen trun}) by iteration (\ref{algor: eigen1})-(\ref{algor: eigen2}), and use the Cauchy criterion to stop the algorithm, i.e., we take $\psi_E\approx\psi_E^{(N)}$ for some $N$ when
$$\left\|\psi_E^{(N)}-\psi_E^{(N-1)}\right\|_{L^\infty}\leq\eps,$$
with $\eps$ a chosen threshed. For $E\approx E_*$, since the nonlinearity is a weak perturbation, i.e. $|\lambda|\ll1$, so as a natural initial guess, we choose $\psi_E^{(0)}=\psi_0$. We measure the accuracy by using the maximum norm of the error
$$
e_E:=E\psi^{(N)}_{E}-\left[-\Delta+V+\lambda\left(\psi_E^{(N)}\right)^{2m}\right]\psi_E^{(N)}.
$$
Tab. \ref{tab: error E} presents the error $e_E$ and the total number of iterations $N$ that is needed under a chosen threshed $\eps=1E-4$ for different $E$ with the same initial guess $\psi_0$. The profiles of the bound state $\psi_E^{(N)}$ under different $E$ are plotted in Fig. \ref{fig}. The corresponding profiles of $\partial_E\psi_E^{(N)}$ are also given in Fig. \ref{fig}.
Tab. \ref{tab: error eps} shows the error and number of iterations under different threshed $\eps=1E-4$ at a fixed $E=-0.8$.

\begin{table}[t!]
  \caption{The error $e_E$ and number of iterations $N$ at different $E$
  under threshed $\eps=1E-4$ with initial guess $\psi_0$.}\label{tab: error E}
  \vspace*{-10pt}
\begin{center}
\def\temptablewidth{1\textwidth}
{\rule{\temptablewidth}{0.75pt}}
\begin{tabular*}{\temptablewidth}{@{\extracolsep{\fill}}llllll}
                                 & $E=-0.9$         &  $E=-0.7$          & $E=-0.5$         & $E=-0.3$   & $E=-0.1$\\[0.25em]
\hline
$\left\|e_E\right\|_{L^\infty}$  & 1.63E-06	        & 4.89E-05	         & 6.11E-05	        &1.07E-04	 &1.43E-04\\ \hline
N                                & 3	            & 4	                 & 6	            &9	         &20\\
\end{tabular*}
{\rule{\temptablewidth}{0.75pt}}
\end{center}
\end{table}

\begin{table}[t!]
  \caption{The error $e_E$ and number of iterations $N$ under different threshed $\eps$
  at $E=-0.8$.}\label{tab: error eps}
  \vspace*{-10pt}
\begin{center}
\def\temptablewidth{1\textwidth}
{\rule{\temptablewidth}{0.75pt}}
\begin{tabular*}{\temptablewidth}{@{\extracolsep{\fill}}lllll}
                                 & $\eps=1E-2$      &  $\eps=1E-3$       & $\eps=1E-4$      & $\eps=1E-5$ \\[0.25em]
\hline
$\left\|e_E\right\|_{L^\infty}$  & 9.32E-04	        & 6.23E-05	         & 2.19E-06	        &2.14E-07\\ \hline
N                                & 2	            & 3	                 & 4	            &5\\
\end{tabular*}
{\rule{\temptablewidth}{0.75pt}}
\end{center}
\end{table}

From Tabs. \ref{tab: error E}-\ref{tab: error eps}, Fig. \ref{fig} and additional results not shown here brevity, we can draw the
following observations:
\begin{enumerate}
\item The iteration algorithm for solving the eigenvalue problem (\ref{eigen trun}) converges. Better initial guess costs less iterations (cf. Tab. \ref{tab: error E}).

\item The algorithm has high accuracy in error $e_E$ and is very efficient. Each step of iteration will make a rapid decay in error $e_E$ (cf. Tab. \ref{tab: error eps}).

\item The nonlinear bound states $\psi_E$ are positive functions fast decaying to zero at far field, and so are the $\partial_E\psi_E$. For the bound states $\psi_E$, the smaller the energy $E$ is, the shaper the solution is (cf. Fig. \ref{fig}).
\end{enumerate}

\begin{remark}
Here in this section, we proposed the algorithm (\ref{algor: eigen1})-(\ref{algor: eigen2}) to solve the eigenvalue problem (\ref{eigen trun}) instead of directly applying the imaginary-time method or Newton-Rapshon  method. We remark that imaginary-time method (also known as the normalized gradient flow method) is applied when the Hamiltonian of the Schr\"{o}dinger equation has discrete spectrum while our problem does not. As for the Newton-Rapshon method, one need an initial guess very close to the root, while in our application this is not always possible. Our proposed algorithm is robust in the choice of the initial guess, although we put some requirements in the convergence theorem as a mathematical technique for the rigorous proof.
\end{remark}

\section{On modulation equations}\label{sec: modulate eqns}
In this section, with the algorithm for the eigenvalue problem in hand, we are going to present the numerical method for solving the modulation equations (\ref{modulation}).

\subsection{Numerical method}
Similar to the eigenvalue problem, since the dispersive wave $\phi$ also decays very fast to zero at far field, we truncate the problem (\ref{mudulated: disper}) onto a finite interval $\Omega=[-L,L]^d$ and impose the homogeneous Dirichlet boundary condition for numerical aspects. The truncated initial boundary value problem of the modulation equations read,
\begin{subequations}\label{modulation trun}
\begin{eqnarray}
&i\partial_t\phi(\mathbf{x},t)=\left[-\Delta+V-E(t)+\dot{\gamma}(t)\right]\phi
+\lambda\left|\psi_{E(t)}+\phi\right|^{2m}\left(\psi_{E(t)}+\phi\right)\label{mudulated trun: disper}\\
&\qquad\qquad\quad\,-\lambda\psi_{E(t)}^{2m+1}+\dot{\gamma}(t)\psi_{E(t)}-i\dot{E}(t)\partial_E\psi_{E(t)},\quad \mathbf{x}\in\Omega,\ \, t>0,\nonumber\\
&\dot{\gamma}(t)=-\frac{\mathfrak{R}\left<F_2,\psi_{E_0}\right>}{\left<\psi_E,\psi_{E_0}\right>},\quad t>0,\\
&\dot{E}(t)=-\frac{\mathfrak{I}\left<F_2,\psi_{E_0}\right>}{\left<\partial_E\psi_E,\psi_{E_0}\right>},\quad t>0,\\
&\gamma(0)=\gamma_0,\quad E(0)=E_0,\quad \phi(\mathbf{x},0)=\phi_0(\mathbf{x}),\quad \mathbf{x}\in\Omega,\label{mudulated trun: ini}\\
&\phi(\mathbf{x},t)=0,\qquad \mathbf{x}\in\partial\Omega.\label{mudulated trun: bd}
\end{eqnarray}
\end{subequations}

Choose the time step size $\tau=\Delta t>0$ and denote the time steps by
$t_n:=n\tau, \quad n=0,1,\ldots.$ To present the scheme, we denote
$$\phi^n(\mathbf{x})\approx\phi(\mathbf{x},t_n),\ \ \psi_{E}^n\approx\psi_{E(t_n)},\ \
\partial_E\psi_{E}^n\approx\partial_E\psi_{E(t_n)},\ \ E^n\approx E(t_n),\ \ \gamma^n\approx\gamma(t_n).$$
and introduce the finite difference operator on some grid functions $f^n$,
$$\delta_tf^n:=\frac{f^{n+1}-f^{n-1}}{2\tau},\qquad n=1,2,\ldots.$$
Then a semi-implicit leap-frog finite difference temporal discretization of (\ref{modulation trun}) reads,
\begin{subequations}\label{modulation semi-dis}
\begin{eqnarray}
&i\delta_t\phi^n=\frac{-\Delta+V}{2}(\phi^{n+1}+\phi^{n-1})+(\delta_t\gamma^n-\delta_tE^n)\phi^n-\lambda(\psi_{E}^n)^{2m+1}
\label{modulation semi-dis: disper}\\
&\qquad\quad\,+\lambda\left|\psi_{E}^n+\phi^n\right|^{2m}\left(\psi_{E}^n+\phi^n\right)+\delta_t\gamma^n\psi_{E}^n-i\delta_tE^n\partial_E\psi_{E}^n,\quad \mathbf{x}\in\Omega,\ \, \nonumber\\
&\delta_t\gamma^n=-\frac{\mathfrak{R}\left<F_2^n,\psi_{E_0}\right>}{\left<\psi_E^n,\psi_{E_0}\right>},\quad n=1,2,\ldots,
\label{modulation semi-dis: gamma}\\
&\delta_tE^n=-\frac{\mathfrak{I}\left<F_2^n,\psi_{E_0}\right>}{\left<\partial_E\psi_E^n,\psi_{E_0}\right>},\quad n=1,2,\ldots,
\label{modulation semi-dis: E}
\end{eqnarray}
\end{subequations}
where
$$F_2^n=\lambda\left|\psi_E^n+\phi^n\right|^{2m}\left(\psi_E^n+\phi^n\right)-\lambda(\psi_E^n)^{2m+1}-\lambda\psi_{E_0}^{2m}\phi^n,
\quad n=0,1,\ldots,$$
and initial values
$$\phi^0=\phi_0,\quad E^0=E_0,\quad \gamma^0=\gamma_0.$$
Since (\ref{modulation semi-dis}) is two-level scheme, we also need the starting values at $t=t_1$. In order to get a second order accuracy in temporal  approximations, they are obtained by the Taylor's expansion of the solution and noticing the equations (\ref{modulation trun}) as
\begin{eqnarray*}
&\phi^1=\phi^0-i\tau\left[(-\Delta+V+\delta_t\gamma^0-\delta_tE^0)\phi^{0}
+\lambda\left|\psi_{E_0}+\phi^0\right|^{2m}\left(\psi_{E_0}+\phi^0\right)\right.\\
&\qquad\ \left.-\lambda(\psi_{E_0})^{2m+1}+\delta_t\gamma^0\psi_{E_0}-i\delta_tE^0\partial_E\psi_{E}^0\right],\\
&E^1=E^0+\tau\delta_tE^0,\qquad \gamma^1=\gamma^0+\tau\delta_t\gamma^0,\\
&\delta_t\gamma^0=-\frac{\mathfrak{R}\left<F_2^0,\psi_{E_0}\right>}{\left<\psi_{E_0},\psi_{E_0}\right>},\qquad
\delta_tE^0=-\frac{\mathfrak{I}\left<F_2^0,\psi_{E_0}\right>}{\left<\partial_E\psi_E^0,\psi_{E_0}\right>}.
\end{eqnarray*}
As for the above spatial derivatives, i.e. the Laplacian operator, we use the standard sine pseudospectral discretization. Thus, our numerical method can be referred as the semi-implicit sine pseudospectral (SISP) method.
Here, $\psi_E^n$ is obtained by algorithm (\ref{algor: eigen1})-(\ref{algor: eigen2}) from (\ref{eigen trun}) and $\partial_E\psi_E^n$ is given by (\ref{dE psi}).

The SISP method is clearly time symmetric. In the scheme of SISP, (\ref{modulation semi-dis: gamma})-(\ref{modulation semi-dis: E}) is fully explicit, while (\ref{modulation semi-dis: disper}) is semi-implicit. So at each time level $t=t_n$, we apply a linear solver, for example the Gauss-Seidel method \cite{Golub}, to get $\phi^{n+1}$.  We remark that here the reason why we put a time average on the operator $-\Delta+V$ in (\ref{modulation semi-dis: disper}) is to get rid of the stability problems \cite{Bao1}.

\begin{remark}
We remark that in the algorithm (\ref{algor: eigen1})-(\ref{algor: eigen2}) for the eigenvalue problem (\ref{eigen trun}) in Section \ref{sec: eigen pro} and the temporally discretized modulation equations (\ref{modulation semi-dis}), one can also use the finite difference method for spatial discretizations. Here we choose the sine pseudospectral method for a high accuracy purpose in the case of zero boundary conditions (\ref{eigen trun b}) and (\ref{mudulated trun: bd}). Corresponding Fourier/cosine pseudospectral discretizations can be applied in the case of periodic/Neumann boundary conditions.
\end{remark}

\subsection{Numerical results}
For simplicity, we also consider the one-dimensional case to present the numerical results by using the SISP method (\ref{modulation semi-dis}) to solve the modulation equations (\ref{modulation trun}).
Choose the same example used in Section \ref{subsec: eigen numres}, i.e. in (\ref{modulation trun}) take
\begin{equation}\label{num: ini0}
V(x)=-2b^2\sech^2(bx),\quad b=m=1,\quad \lambda=0.1,\quad \Omega=[-20,20],
\end{equation}
and take the initial conditions as
\begin{equation}\label{num: ini}
E_0=-0.8,\quad \gamma_0=0.5,\quad \phi_0(x)=5xe^{-2x^2}\cdot\psi_{E_0}(x).
\end{equation}
We remark here the chosen $\psi_{E_0}$ and $\phi_0$ satisfy the orthogonal condition in (\ref{ansatz: orth}), since $\psi_{E_0}(x)$ is even.

\vspace{1mm}
\emph{Accuracy test}
\vspace{1mm}

First of all, we test the correctness and accuracy of the proposed SISP method. To show the modulation equations with SISP solve the NLS equation (\ref{nls}) correctly, we first solve the modulation equations (\ref{modulation trun}) numerically by the SISP (\ref{modulation semi-dis}) to get $\phi^M(x),$ $\gamma^M$ and $E^n$ for $0\leq n\leq M=T/\tau$, and use the decomposition (A) to construct the numerical solutions $\Phi^M(x)$ of the NLS (\ref{nls}), i.e.
\begin{equation}\label{Phi_num}
\Phi^M(x)=\fe^{-i\left(\mathcal{E}^M-\gamma^M\right)}\left[\psi_{E^M}(x)+\phi^M(x)\right],\quad x\in \Omega,
\end{equation}
where we apply the trapezoidal rule to approximate the integration of $E$ in time in (A2), i.e.
$$\mathcal{E}^M:=\frac{\tau}{2}\sum_{n=0}^{M-1}(E^n+E^{n+1})\approx\int_0^TE(s)\ud s.$$
Then, we compute the error
\begin{equation}\label{e_Phi}
e_{\Phi}(x,T):=\Phi(x,T)-\Phi^M(x),\quad x\in\Omega,
\end{equation}
where the exact solution $\Phi(x,T)$ of the NLS is obtained by classical numerical methods such as the time-splitting sine spectral method \cite{Bao1,BaoMarkowich} with very small step size, e.g. $\tau=10^{-4},\ h=1/16$. Meanwhile, we test the convergence of the SISP method in solving the three decomposed components $(E,\gamma,\phi)$, i.e. we compute the error
\begin{equation}\label{e_three}
e_E:=E^M-E(T),\quad e_\gamma:=\gamma^M-\gamma(T),\quad e_\phi(x,T):=\phi^M-\phi(x,T).
\end{equation}
Here for the `exact' $E(T),\ \gamma(T),\ \phi(x,T)$, we use the SISP method with very small time step and mesh size, e.g. $\tau=10^{-4},\ h=1/16$.
We test the temporal and spatial discretization errors of the SISP method separately. Firstly, for the discretization error in time, we take a fine mesh size $h=1/8$ such that the error from the discretization in space is negligible compared to the temporal discretization error. The errors (\ref{e_Phi})-(\ref{e_three}) under maximum norm are presented at $T=0.5$ and tabulated in Tab. \ref{tab: error temp}. Secondly, for the discretization error in space, we take a very small time step $\tau=10^{-4}$  such that the error from the discretization in time is negligible compared to the spatial discretization error. The corresponding errors under maximum norm are presented at $T=0.5$ as well and tabulated in Tab. \ref{tab: error spat}.

\vspace{1mm}
\emph{Conservation law test}
\vspace{1mm}

\begin{figure}[t!]
{
$$
\begin{array}{cc}
\psfig{figure=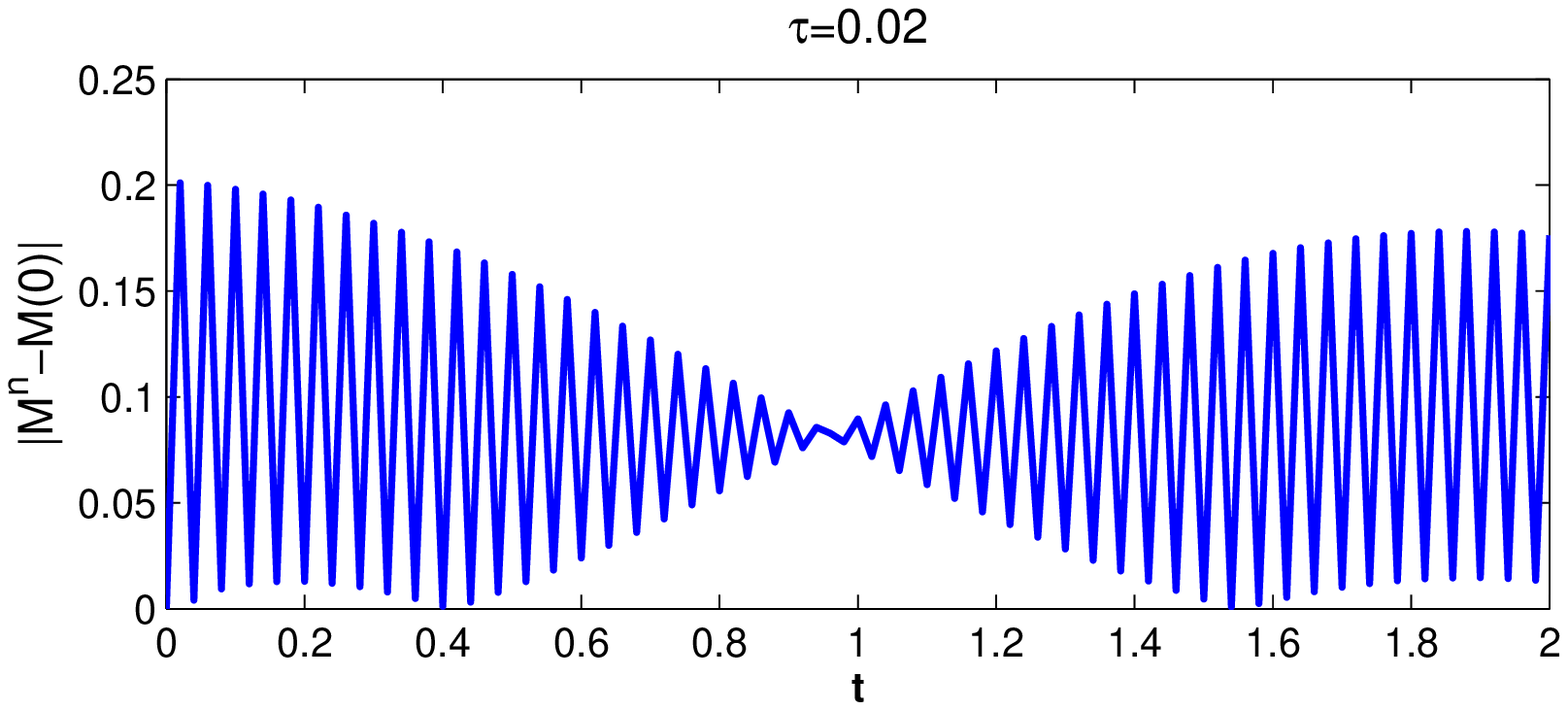,height=5cm,width=6cm}&\psfig{figure=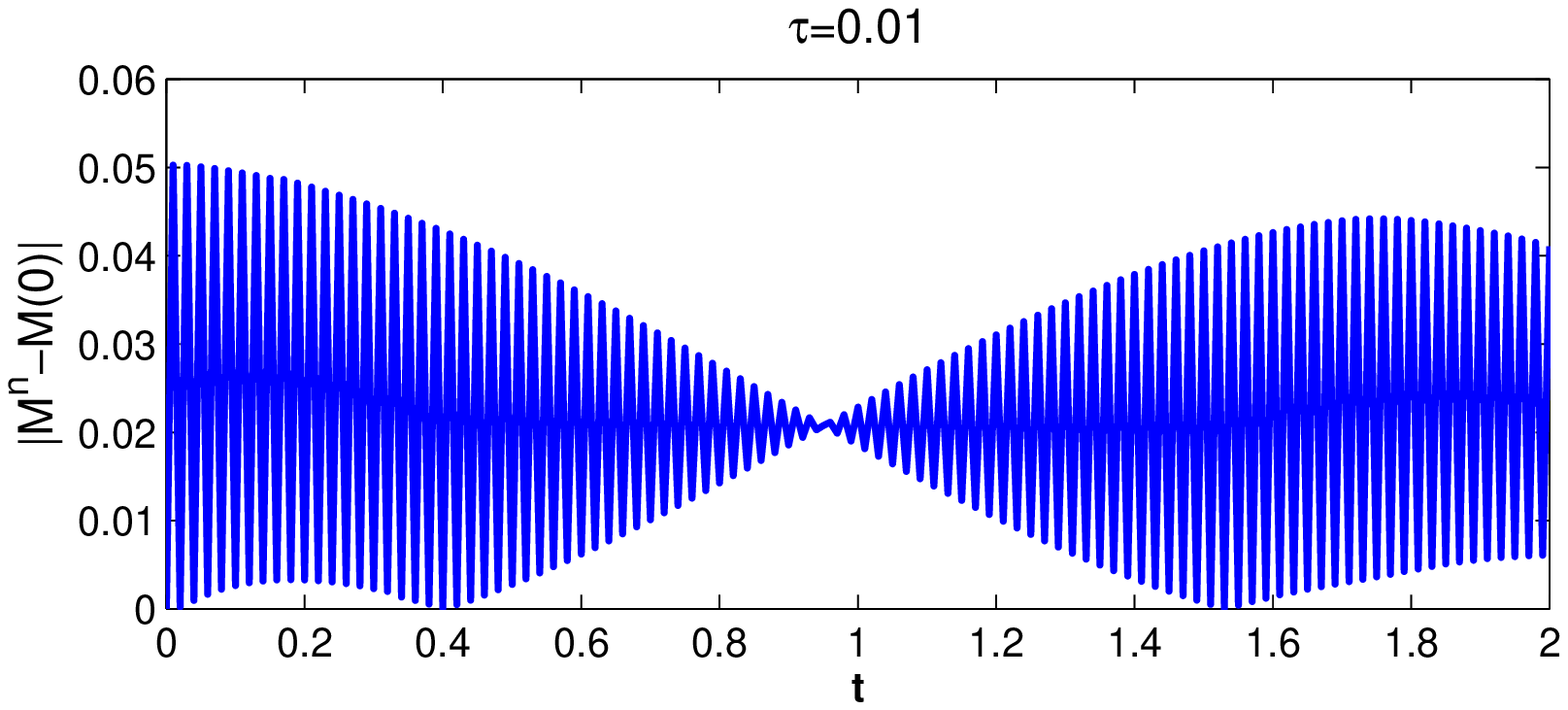,height=5cm,width=6cm}
\end{array}
$$
}
\caption{Mass fluctuations $\left|M^n-M(0)\right|$ of the NLS during the computation of SISP under different time step $\tau$ with mesh size $h=1/16$. }\label{fig:M}
\end{figure}

\begin{figure}[t!]
{$$
\begin{array}{cc}\psfig{figure=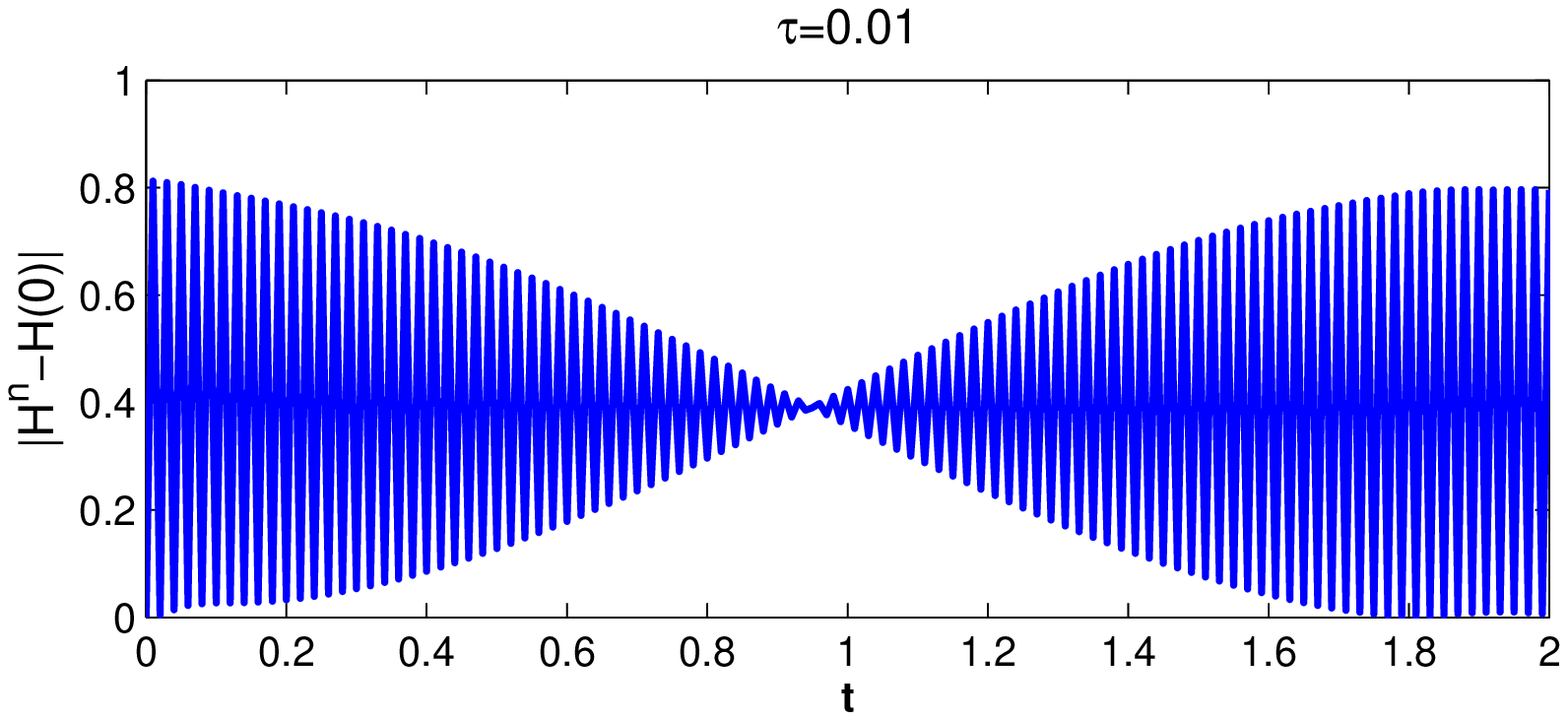,height=5cm,width=6cm}&
\psfig{figure=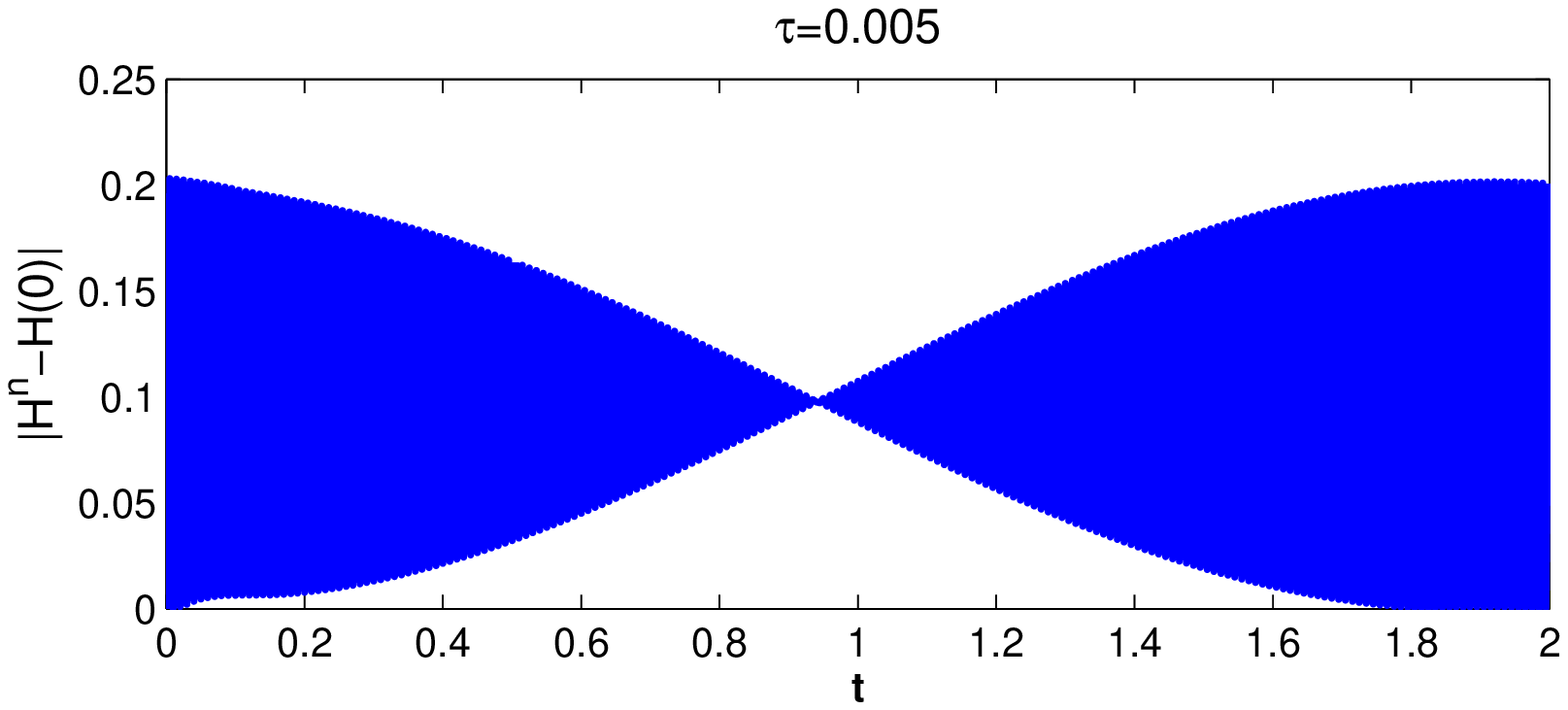,height=5cm,width=6cm}\end{array}
$$
}
\caption{Energy fluctuations $\left|H^n-H(0)\right|$ of the NLS during the computation of SISP under different time step $\tau$ with mesh size $h=1/16$. }\label{fig:H}
\end{figure}

Moreover, with the numerical solutions of the NLS (\ref{nls}) from the SISP method (\ref{modulation semi-dis}), i.e. $\Phi^n(x)$ constructed similarly as in (\ref{Phi_num}) for $n=0,1,\ldots,$ we compute the numerical mass $M^n$ and Hamiltonian $H^n$:
\begin{eqnarray*}
M^n&:=\int_{\bR^d}\left|\Phi^n(x)\right|^2\ud x,\\
H^n&:=\int_{\bR^d}\left[\left|\frac{d}{dx}\Phi^n(x)\right|^2+V(x)\left|\Phi^n(x)\right|^2
+\frac{\lambda}{m+1}\left|\Phi^n(x)\right|^{2m+2}
\right]\ud x,
\end{eqnarray*}
with sine pseudospectral discretization for spatial derivatives and integrations in above. The fluctuations in the numerical mass and Hamiltonian, i.e.
$|M^n-M(0)|$ and $|H^n-H(0)|$ for $n=0,1,\ldots$, during the computations of the SISP method are plotted in Fig \ref{fig:M} and Fig \ref{fig:H}, respectively, with a small mesh size $h=1/16$ under different time steps $\tau$.

\begin{table}[t!]
  \caption{The temporal error and convergence rate of the SISP method for the modulation equation under different time step $\tau$ with $h=1/8$ at $T=0.5$.}\label{tab: error temp}
  \vspace*{-10pt}
\begin{center}
\def\temptablewidth{1\textwidth}
{\rule{\temptablewidth}{0.75pt}}
\begin{tabular*}{\temptablewidth}{@{\extracolsep{\fill}}llllll}
                                 & $\tau_0=0.01$       &  $\tau_0/2$          & $\tau_0/4$           & $\tau_0/8$ \\[0.25em]
\hline
$\left\|e_{\Phi}\right\|_{L^\infty}$  & 9.09E-02	&2.21E-02	       &5.40E-03	        &1.40E-03\\
rate                              & --	            & 2.04	           &2.03	            &1.94\\ \hline
$|e_E|$                           & 7.98E-05	    &1.98E-05	       &4.88E-06	        &1.20E-06\\
rate	                          & --              & 2.01	           & 2.02	            &2.03\\ \hline
$|e_\gamma|$                      & 1.30E-03	    & 3.23E-04	       & 8.01E-05	        &1.99E-05\\
rate                              & --	            & 2.01	           &2.01	            &2.01\\ \hline
$\left\|e_{\phi}\right\|_{L^\infty}$ &9.11E-02	    &2.21E-02	       &5.40E-03	        &1.30E-03\\
rate                              & --	            & 2.04	           & 2.03	            &2.05\\
\end{tabular*}
{\rule{\temptablewidth}{0.75pt}}
\end{center}
\end{table}

\begin{table}[t!]
  \caption{The spatial error of the SISP method for the modulation equation under different mesh size $h$ with $\tau=10^{-4}$ at $T=0.5$.}\label{tab: error spat}
  \vspace*{-10pt}
\begin{center}
\def\temptablewidth{1\textwidth}
{\rule{\temptablewidth}{0.75pt}}
\begin{tabular*}{\temptablewidth}{@{\extracolsep{\fill}}llllll}
                                 & $h=1$            &  $h=1/2$          & $h=1/4$           & $h=1/8$ \\[0.25em]
\hline
$\left\|e_{\Phi}\right\|_{L^\infty}$  &6.95E-01	    &4.57E-02	        &2.33E-04	       &9.68E-06\\ \hline
$|e_E|$                               & 7.90E-03	&2.27E-04	        &2.42E-08	       &1.01E-10\\ \hline
$|e_\gamma|$                          & 1.01E-01	&2.50E-03	        &4.93E-08	       &2.31E-10\\ \hline
$\left\|e_{\phi}\right\|_{L^\infty}$  &7.06E-01	    &4.58E-02	        &2.34E-04	       &9.67E-06\\
\end{tabular*}
{\rule{\temptablewidth}{0.75pt}}
\end{center}
\end{table}

Clearly, from Tabs. \ref{tab: error temp}-\ref{tab:   error spat}, we can conclude that the SISP method (\ref{modulation semi-dis}) solves the multichannel solutions based on the modulation equations for the NLS equation correctly and accurately. The numerical
method has second order accuracy in time and spectral accuracy in space. The two conserved quantities are just small fluctuations from the exact one, and the fluctuations will decrease zero as time step decreases.

\section{Explorations on multichannel solutions and comparisons}

\begin{figure}[t!]
$$
\begin{array}{cc}
\psfig{figure=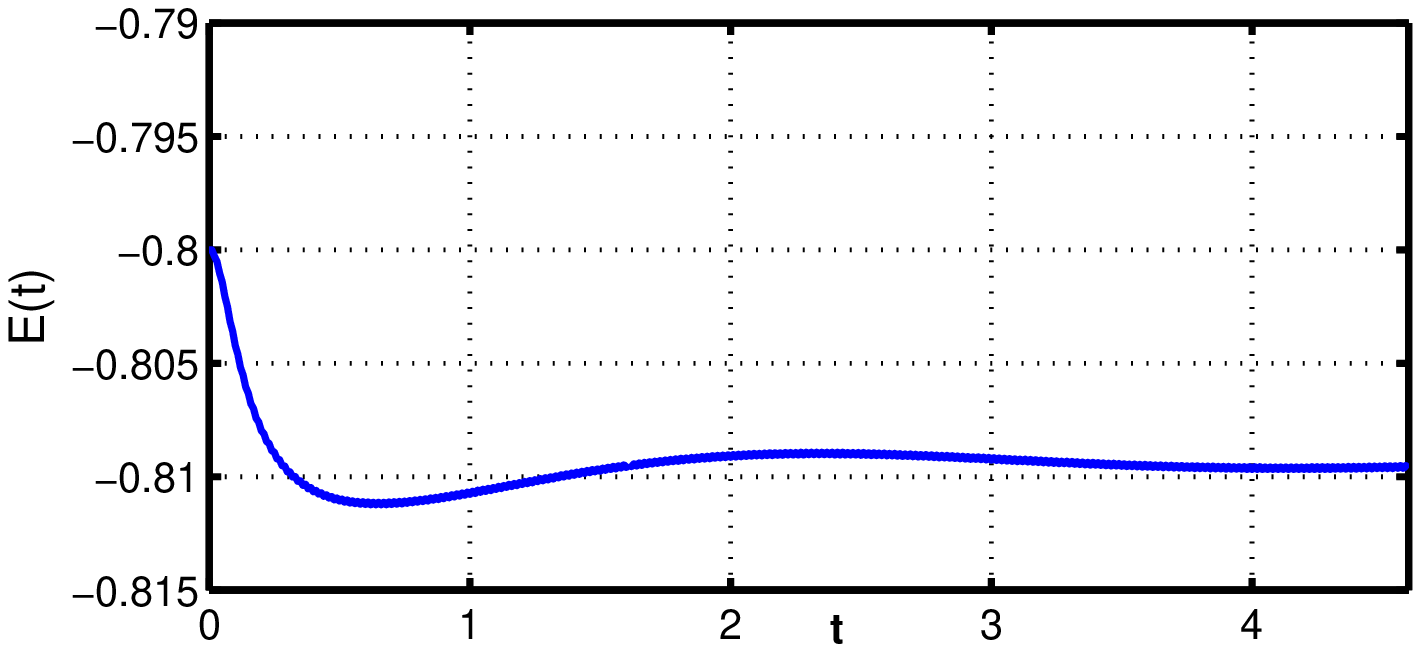,height=5.2cm,width=6.3cm}&
\psfig{figure=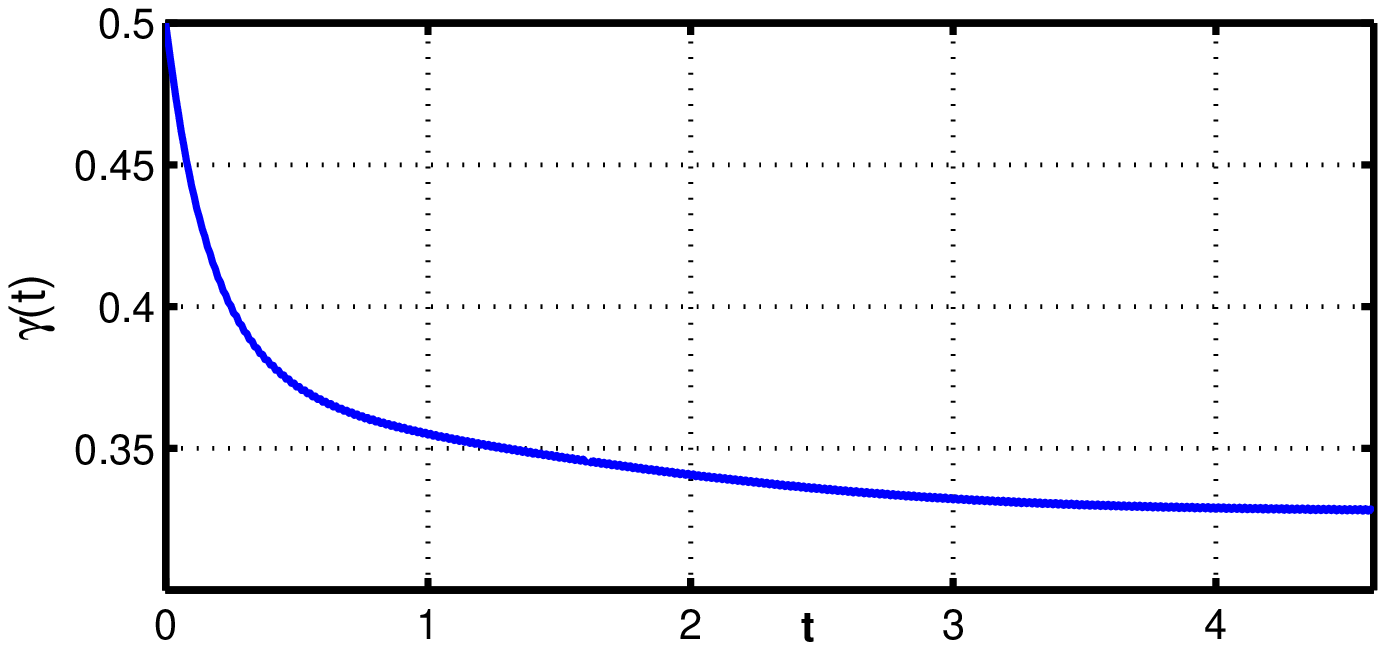,height=5.2cm,width=6.3cm}
\end{array}
$$
\caption{Dynamics of the collective coordinates $E(t)$ and $\gamma(t)$. }\label{fig2}
\end{figure}

\begin{figure}[t!]
{\centerline{
\psfig{figure=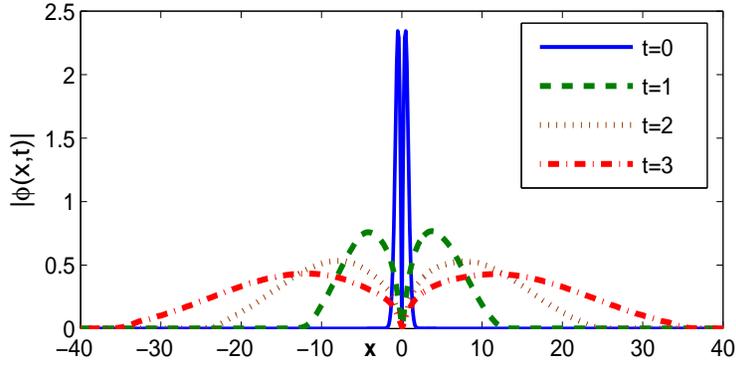,height=5.2cm,width=11cm}}}
\caption{Profiles of the dispersive wave $|\phi(x,t)|$ at different $t$. }\label{fig3}
\end{figure}

\begin{figure}
$$
\begin{array}{cc}
\psfig{figure=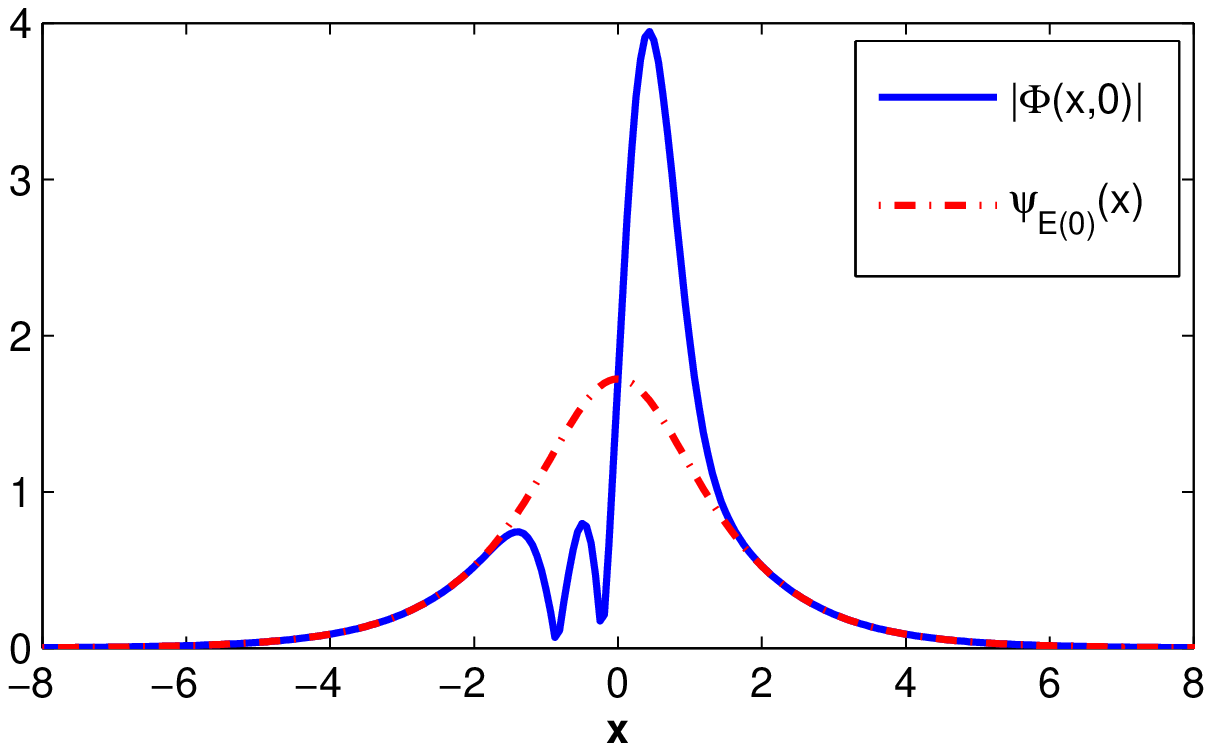,height=4.8cm,width=6.3cm}&\psfig{figure=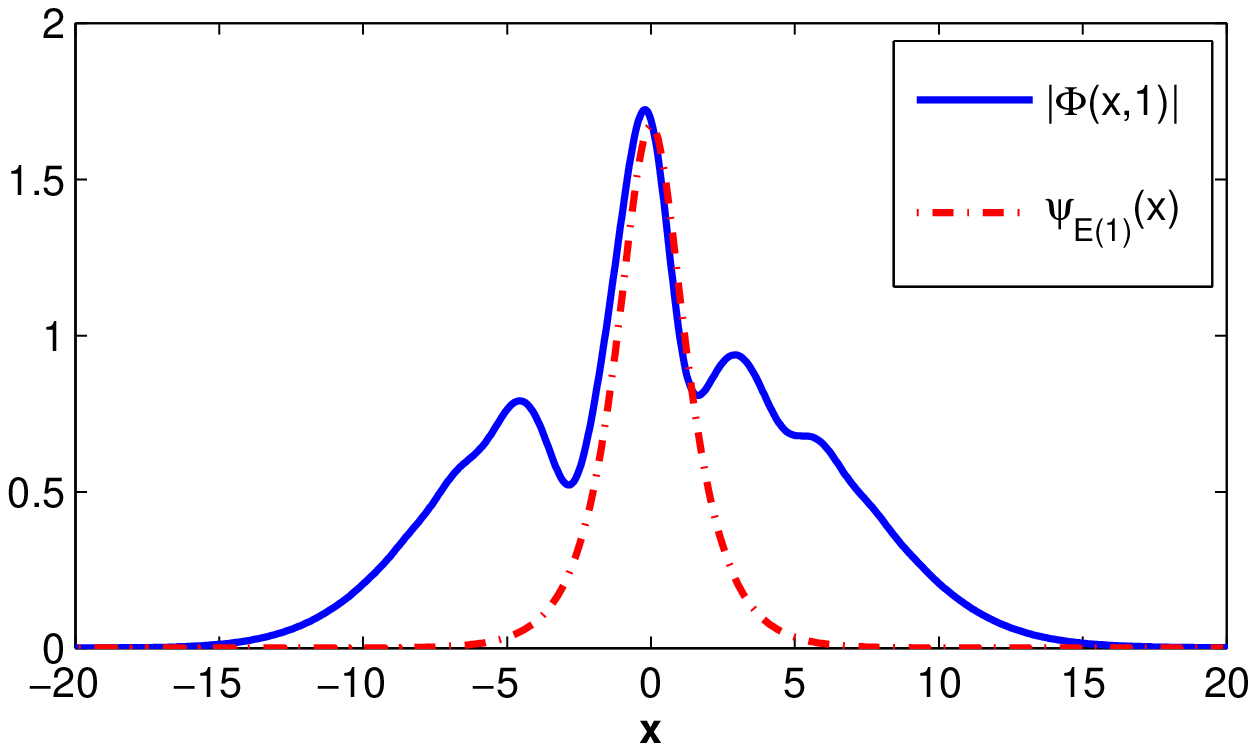,height=4.8cm,width=6.3cm}\\
\psfig{figure=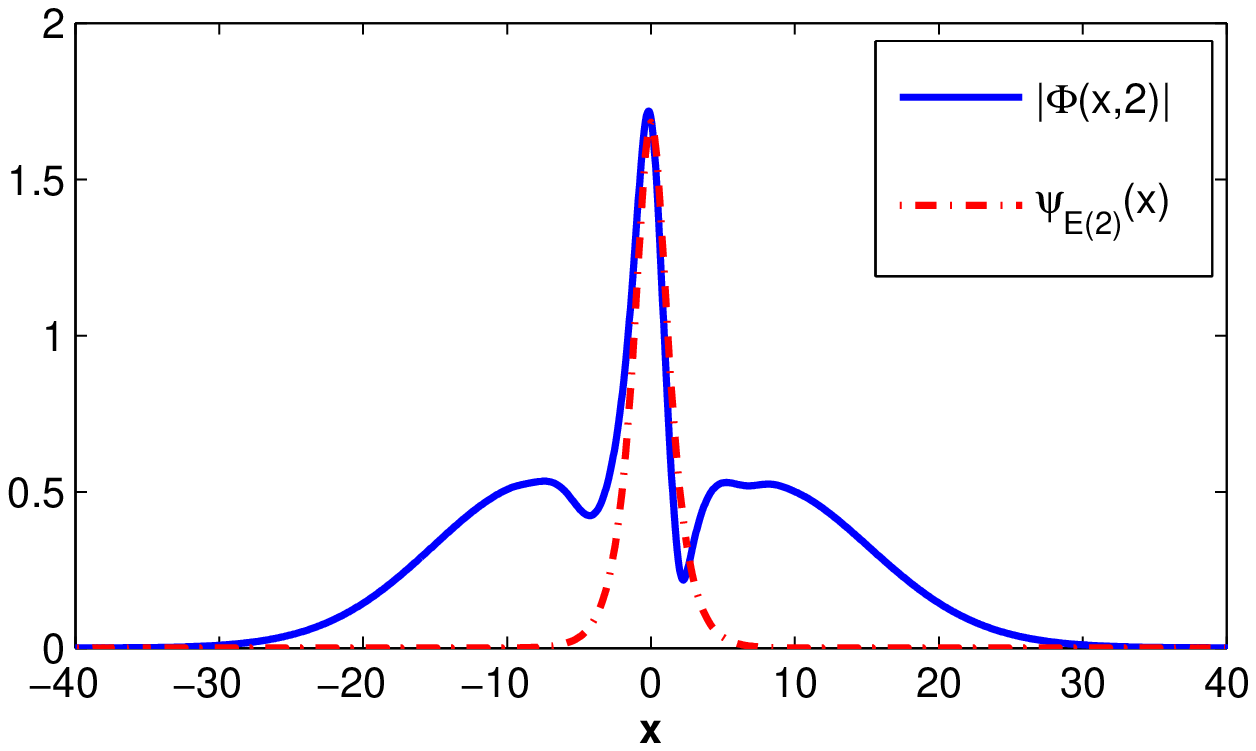,height=4.8cm,width=6.3cm}&\psfig{figure=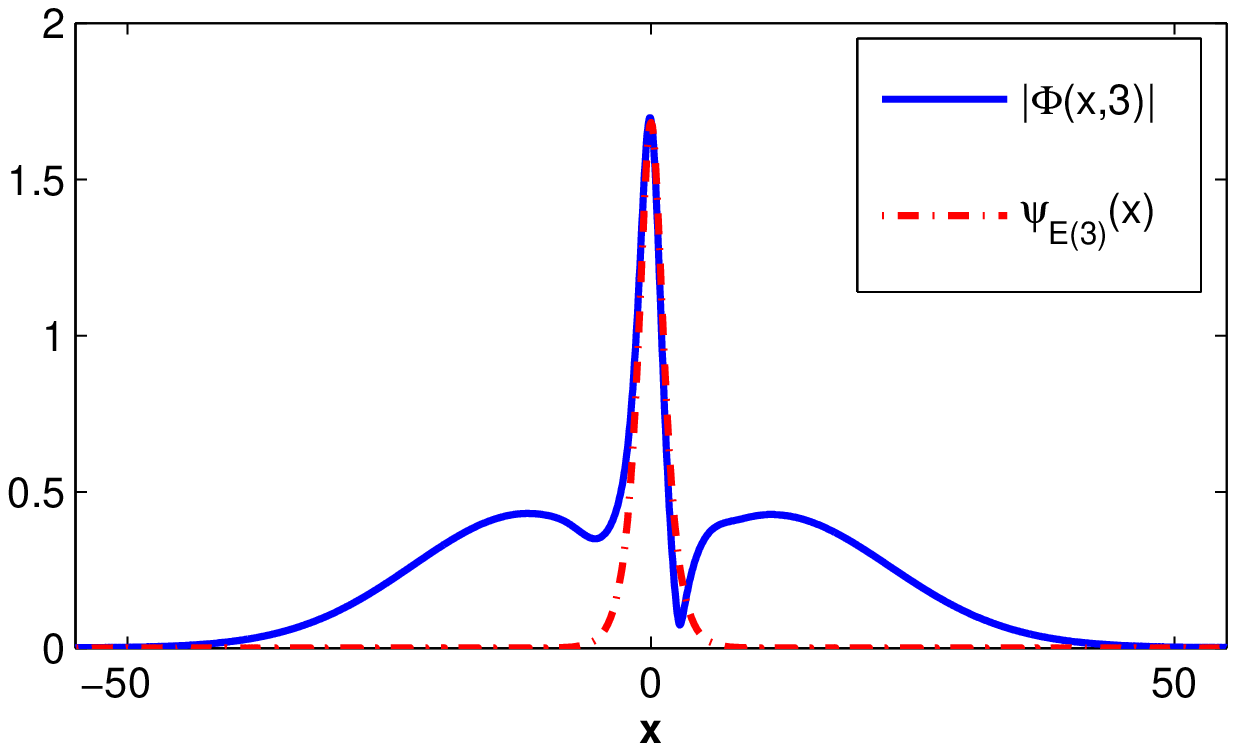,height=4.8cm,width=6.3cm}
\end{array}
$$
\caption{Solution of the NLS $|\Phi(x,t)|$ and the soliton $\psi_{E(t)}(x)$ at different $t$. }\label{fig2.5}
\end{figure}

Now we apply the SISP method to study the dynamics of the multichannel solutions with setup (\ref{num: ini0}) and (\ref{num: ini}) numerically. In order to provide a long time accurate simulation, we choose a large domain $\Omega=[-60,60]$, such that the dispersive wave $\phi$ is always away from the zero boundary during the computation. Taking step size $\tau=1E-3,\ h=1/8$, we solve the modulation equation (\ref{modulation trun}) till the collective coordinates $E(t)$ and $\gamma(t)$ reach the steady state. The dynamics of the collective coordinates $E(t)$ and $\gamma(t)$ are shown in Fig. \ref{fig2}. The profiles of the dispersive wave $\phi(x,t)$ at different time are shown in Fig. \ref{fig3}. The dynamics of the original solution of the NLS $\Phi(x,t)$  and the solition $\psi_{E(t)}(x)$ are plotted in Fig. \ref{fig2.5}.

Based on Figs. \ref{fig2}-\ref{fig2.5}, we can draw the
following observations:
\begin{enumerate}
\item The collective coordinates $E(t)$ and $\gamma(t)$ converge to the steady state as $t$ goes to infinity (cf. Fig. \ref{fig2}), and the dispersive part $\phi(x,t)$ spreads out to far field (cf. Fig. \ref{fig3}). The numerical results match with theoretical results in \cite{Soffer1} very well.

\item The function $E(t)$ is not monotone in time $t$ (cf. the left figure in Fig. \ref{fig2}). This indicates that the process of the dynamics of the soliton and the dispersive wave is not monotone.

\item The dispersive wave $\phi$ in this case has a large expanding velocity (cf. Fig. \ref{fig3}). It is because that the chosen initial perturbation, i.e. the $\phi_0$ in (\ref{num: ini}), has a large $H^1$-norm. For the kind of situation, we remark that using the absorbing boundary conditions could be a more efficient way of study, which will be done in future consequential work.

\item For small time, the radiation term makes a big difference between the solution of NLS and the soliton (cf. Fig. \ref{fig2.5}). Thus, approximating the multichannel solution by the collective coordinates method in this example will call for a significant radiation correction. For large time, as the waves get close to the boundary of the computational domain, traditional numerical methods working the NLS cannot tell the soliton from the dispersive wave, while our approach of the modulation equations distinguish them exactly from the beginning.
\end{enumerate}

\emph{Comparisons with classical numerical solver}

\begin{figure}[t!]
$$
\begin{array}{cc}
\psfig{figure=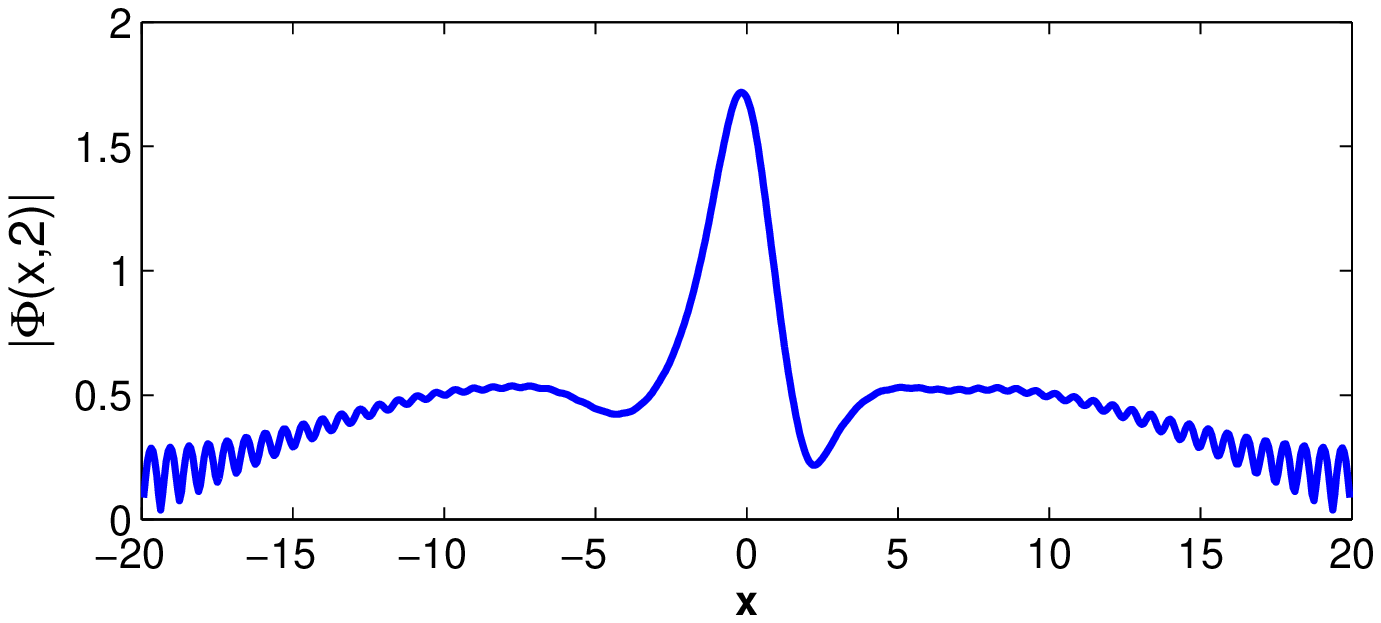,height=4.8cm,width=6.3cm}&\psfig{figure=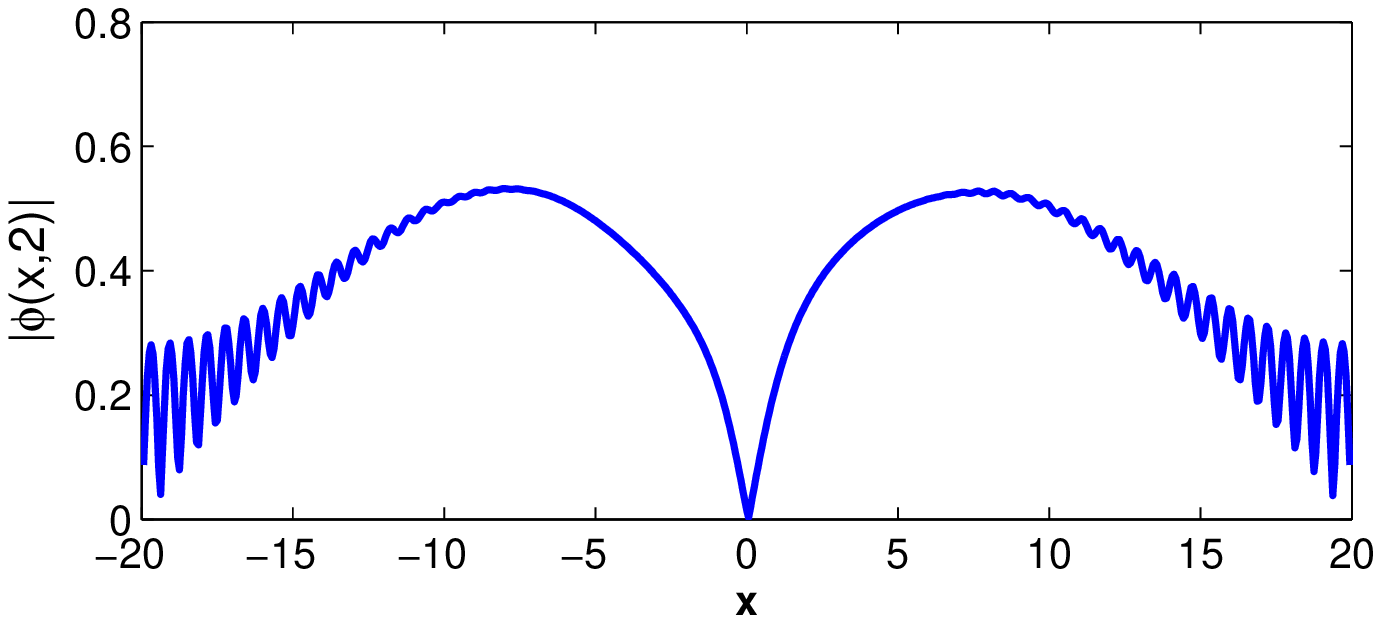,height=4.8cm,width=6.3cm}\\
\psfig{figure=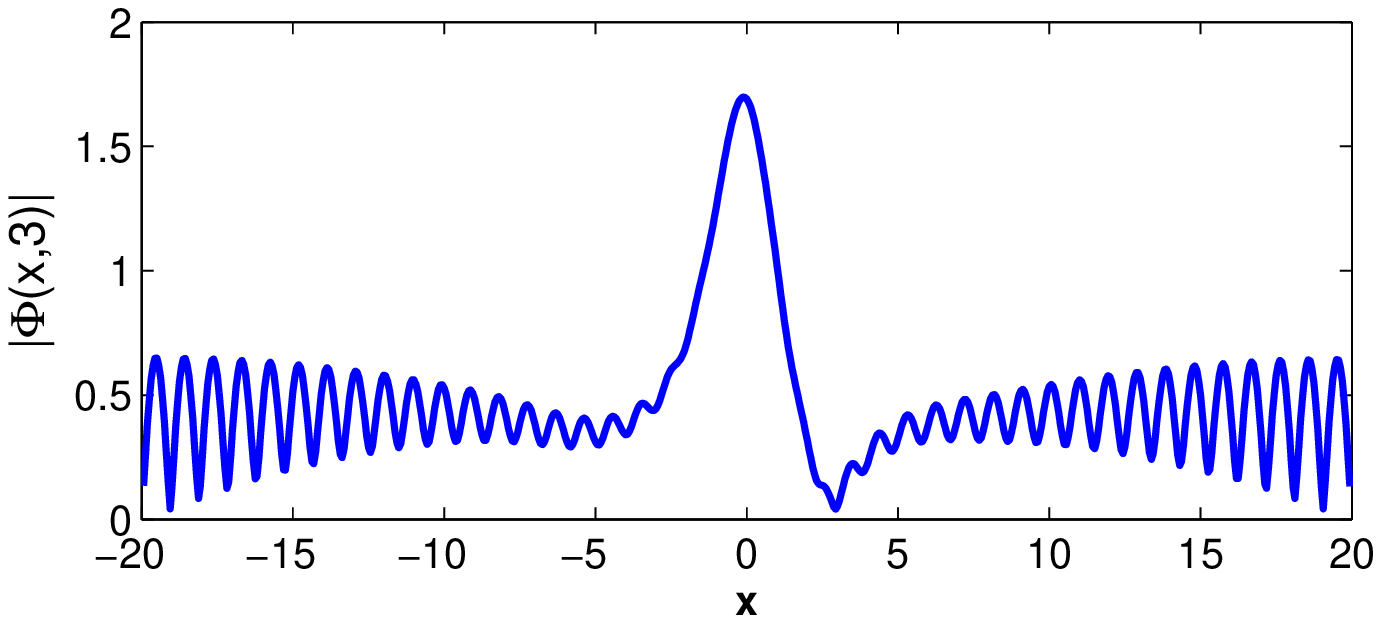,height=4.8cm,width=6.3cm}&\psfig{figure=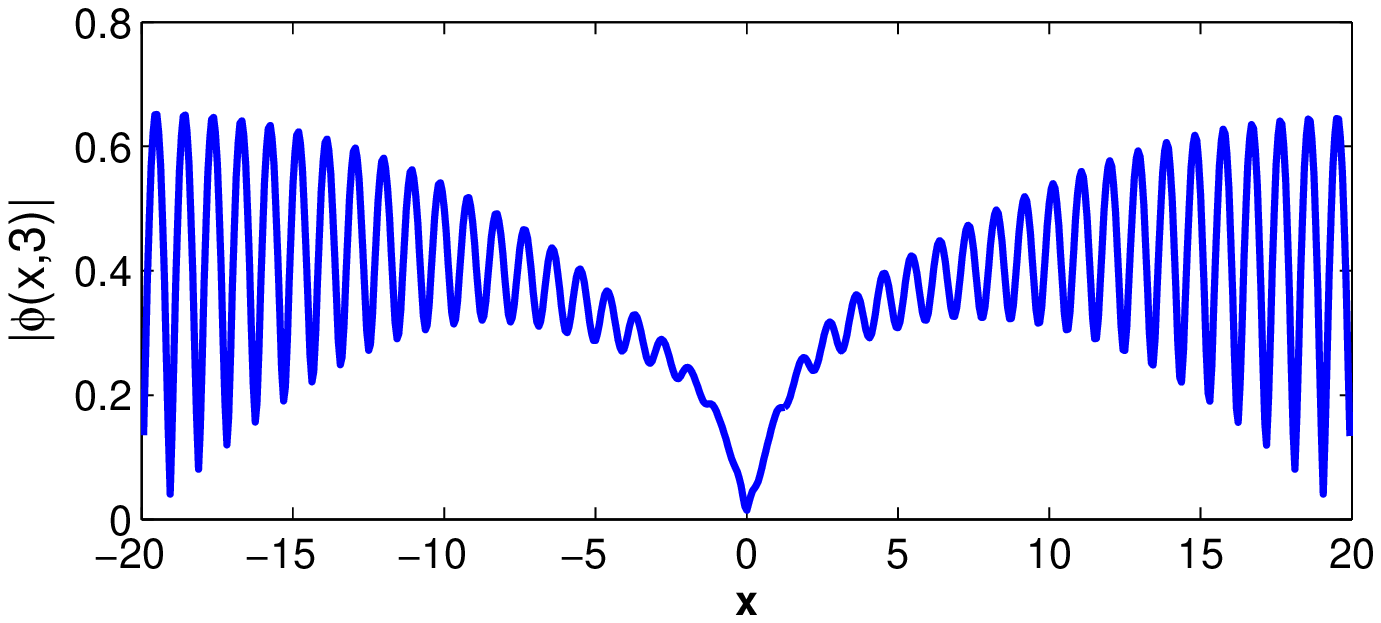,height=4.8cm,width=6.3cm}\\
\psfig{figure=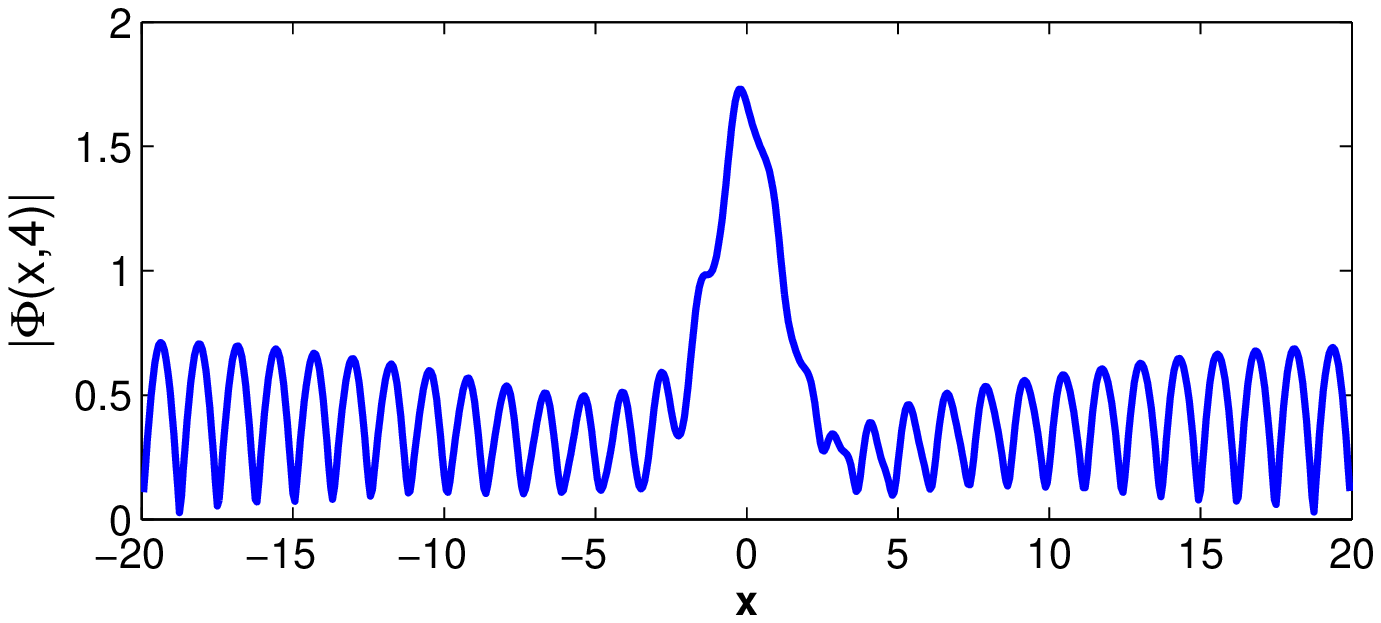,height=4.8cm,width=6.3cm}&\psfig{figure=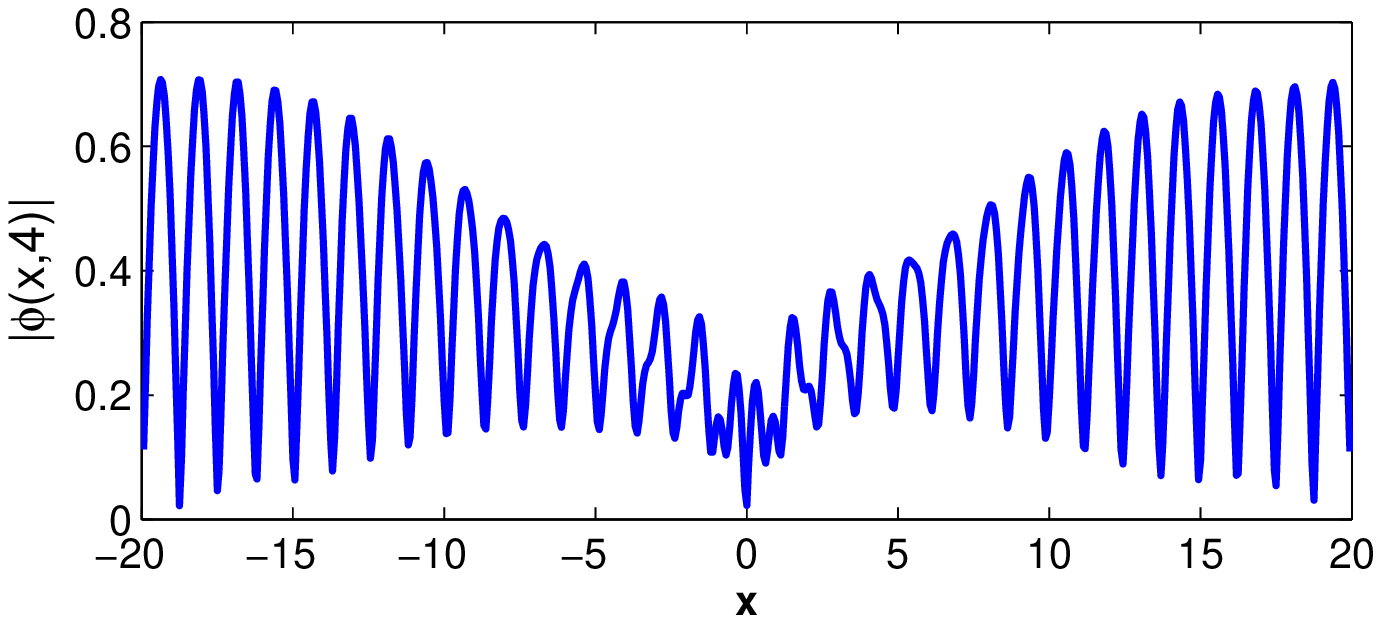,height=4.8cm,width=6.3cm}
\end{array}
$$
\caption{Numerical solutions of the NLS $|\Phi(x,t)|$ from classical solver and the dispersive wave $|\phi(x,t)|$ from modulation equations method on domain $\Omega=[-20,20]$ at different $t$. }\label{figadd}
\end{figure}

To further show the advantages of modulation equations approach over classical numerical methods, we shorten the chosen computational domain to $\Omega=[-20,20]$, such that for time $t\geq2$, the outgoing dispersive waves already reach the boundaries and are reflected back by the zero boundary conditions. Again we solve the NLS via the modulation equations approach with the SISP method with small time step and mesh size till $t=4$. As comparisons, we also solve the NLS directly with classical numerical method like the time-splitting sine spectral method or the finite difference time domain methods. Fig. \ref{figadd} shows the numerical solutions of the NLS $|\Phi(x,t)|$ from the classical solver and the dispersive wave $|\phi(x,t)|$ from the modulation equations method at different $t$ after reflections happen. To check the influence bring by the reflected dispersive waves to the collective coordinates in our modulation approach, we compare the numerical values of $E(t)$ and $\gamma(t)$ obtained from the modulation equations method with the accurate values from Fig. \ref{fig2}. The errors at different $t$ are shown in Tab. \ref{tabadd}.
Based on Tab. \ref{tabadd} and Fig. \ref{figadd}, we can see: Both the numerical solution of the NLS from the classical solver and the numerical solution of the dispersive wave from modulation equations method are destroyed by the reflections in the computational domain compared to the exact solutions from Fig. \ref{fig2.5}. However in modulation equations approach, the error caused by the reflections to $E(t)$ and $\gamma(t)$ is always small during the computation. That is to say the small computational domain and the imposed boundary conditions will only break the dispersive part of the solution, but will hardly affect the soliton. The observation also give some clues to believe that using an absorbing boundary condition to the dispersive wave in (\ref{mudulated trun: disper}) could offer better results. While in classical numerical solvers, reflections will break both the soliton part and the dispersive part in the solution. Again we note that applying absorbing boundaries directly to the NLS will drag out the solitons as well \cite{SStu}. Thus, to isolate the solitons from some initial matter waves in physical applications as BEC or optical lattice \cite{BBB,Soffer1}, classical solvers need to choose a large computational domain and simulate for large time which is quite expensive in computational cost, while our approach do not.

\begin{table}[t!]
  \caption{Error in collective coordinates $E(t)$ and $\gamma(t)$ caused by the reflections of the dispersive wave at the boundaries on domain $\Omega=[-20,20]$.}\label{tabadd}
  \vspace*{-10pt}
\begin{center}
\def\temptablewidth{1\textwidth}
{\rule{\temptablewidth}{0.75pt}}
\begin{tabular*}{\temptablewidth}{@{\extracolsep{\fill}}lllll}
                                 & $t=2$            &  $t=3$&  $t=4$\\[0.25em]
\hline
$|e_E|$                               & 5.88E-05	&5.92E-05	&6.56E-05   \\ \hline
$|e_\gamma|$                          & 1.00E-03	&1.00E-03	&4.29E-04        \\
\end{tabular*}
{\rule{\temptablewidth}{0.75pt}}
\end{center}
\end{table}

\emph{Comparisons with collective coordinates method}

\begin{figure}[t!]
$$
\begin{array}{cc}
\psfig{figure=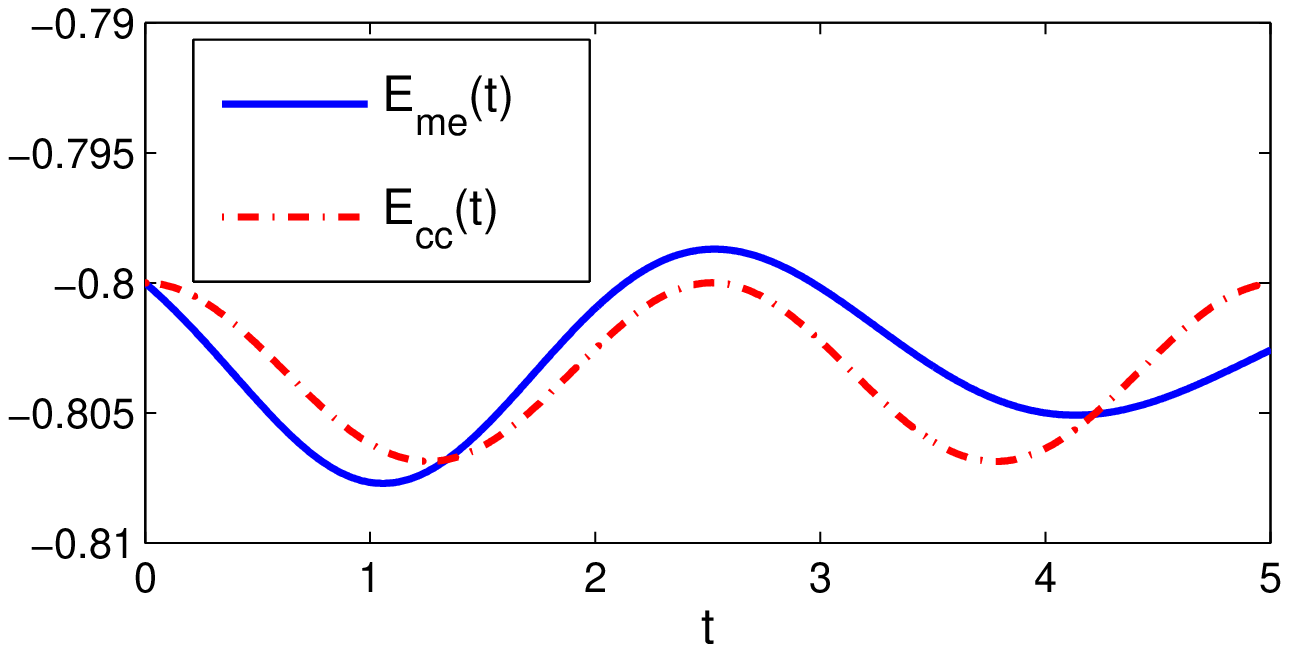,height=4.8cm,width=6.3cm}&\psfig{figure=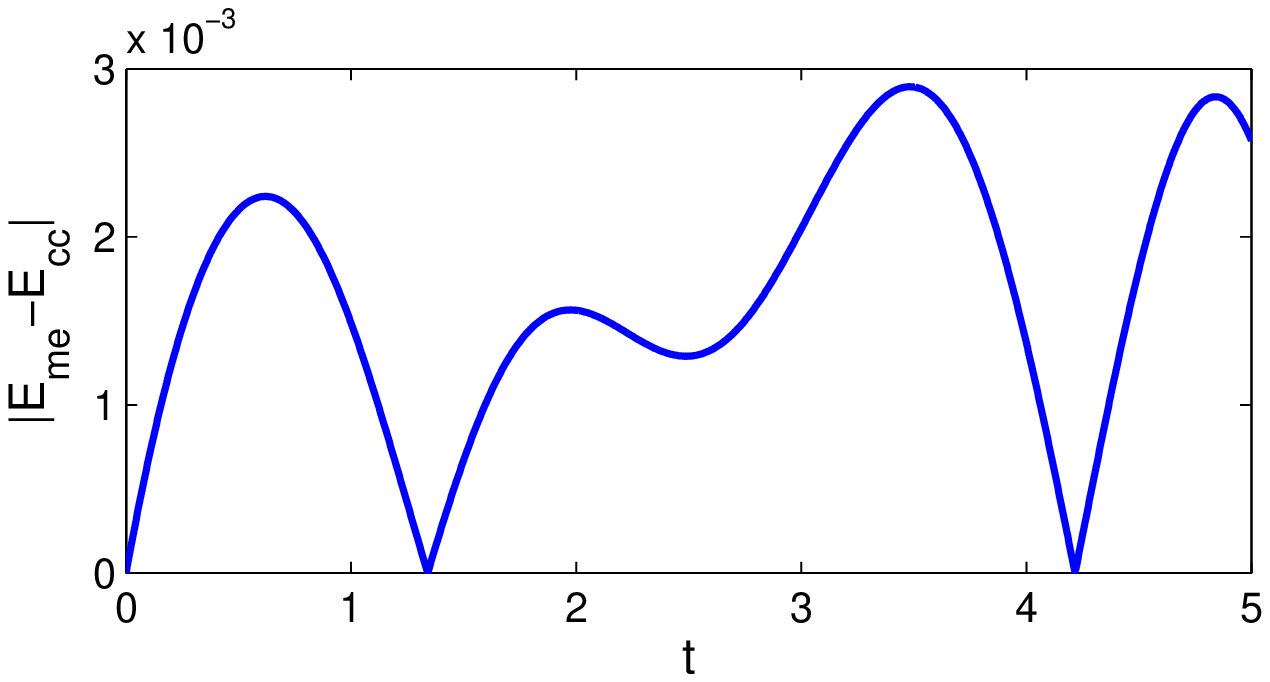,height=4.8cm,width=6.3cm}\\
\psfig{figure=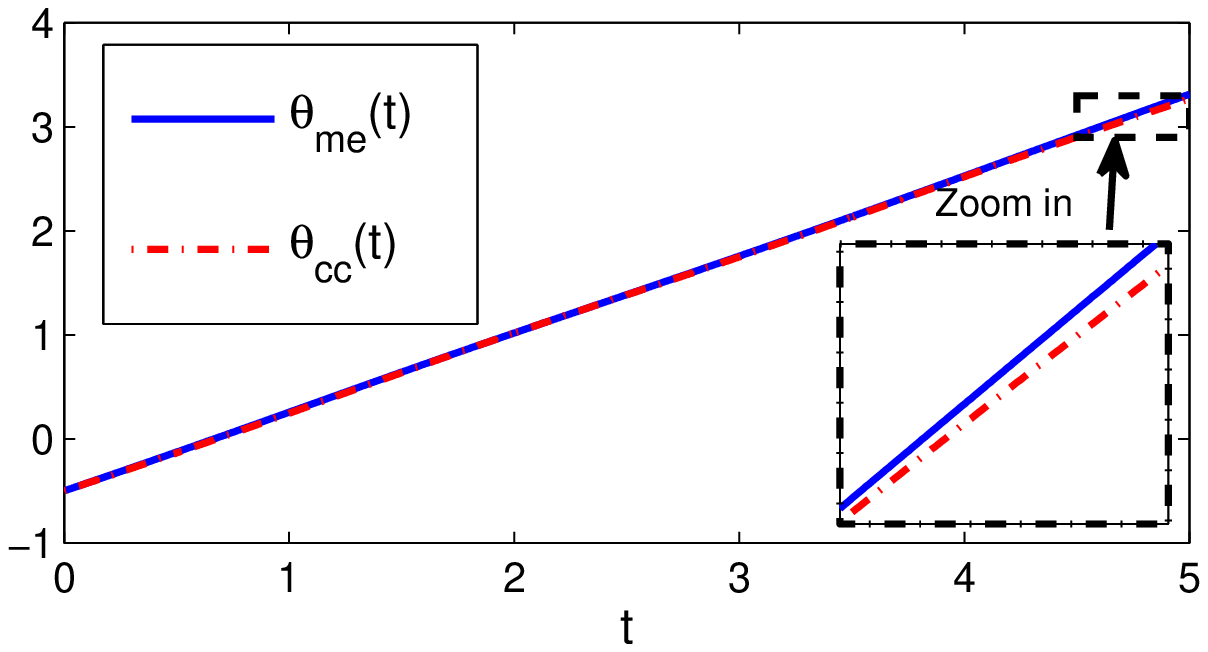,height=4.8cm,width=6.3cm}&\psfig{figure=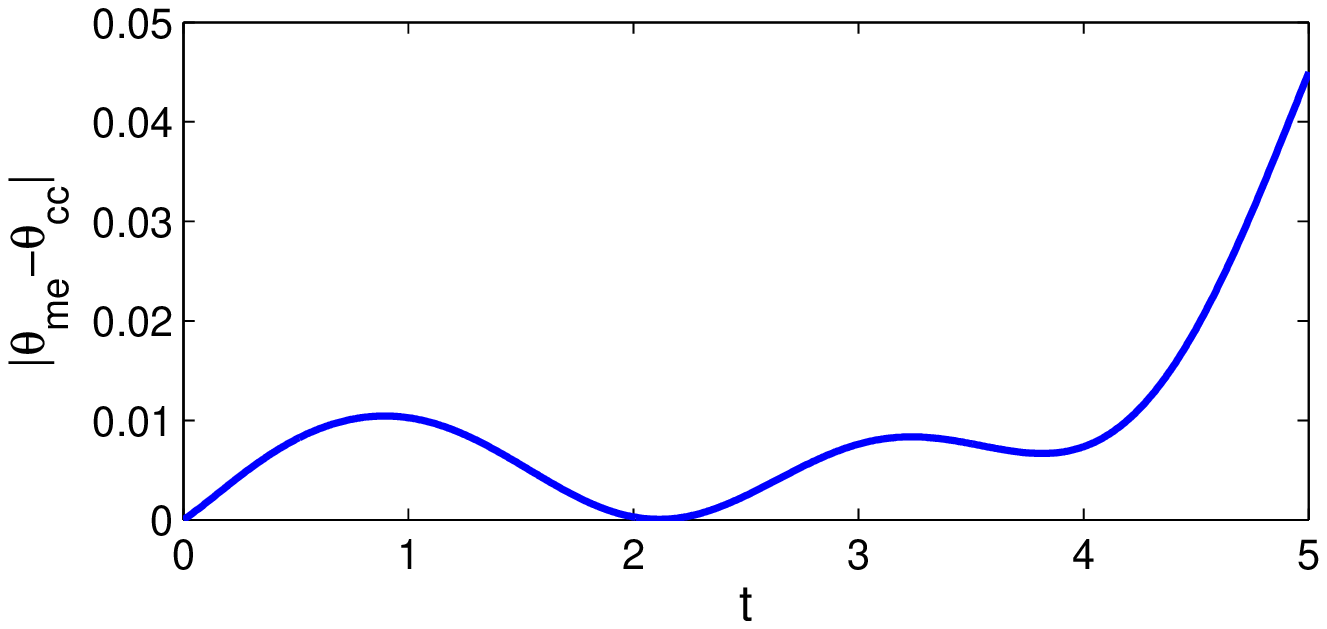,height=4.8cm,width=6.3cm}
\end{array}
$$
\caption{Left column: dynamics of the $E_{me}(t),\theta_{me}(t)$ given by modulation equation method and the $E_{cc}(t),\theta_{cc}(t)$ given by collective coordinates method in the comparison example (\ref{model}). Right column: difference $|E_{me}(t)-E_{cc}(t)|$ and $|\theta_{me}(t)-\theta_{cc}(t)|$ between the two methods.}\label{figadd1}
\end{figure}

\begin{figure}[t!]
$$
\begin{array}{cc}
\psfig{figure=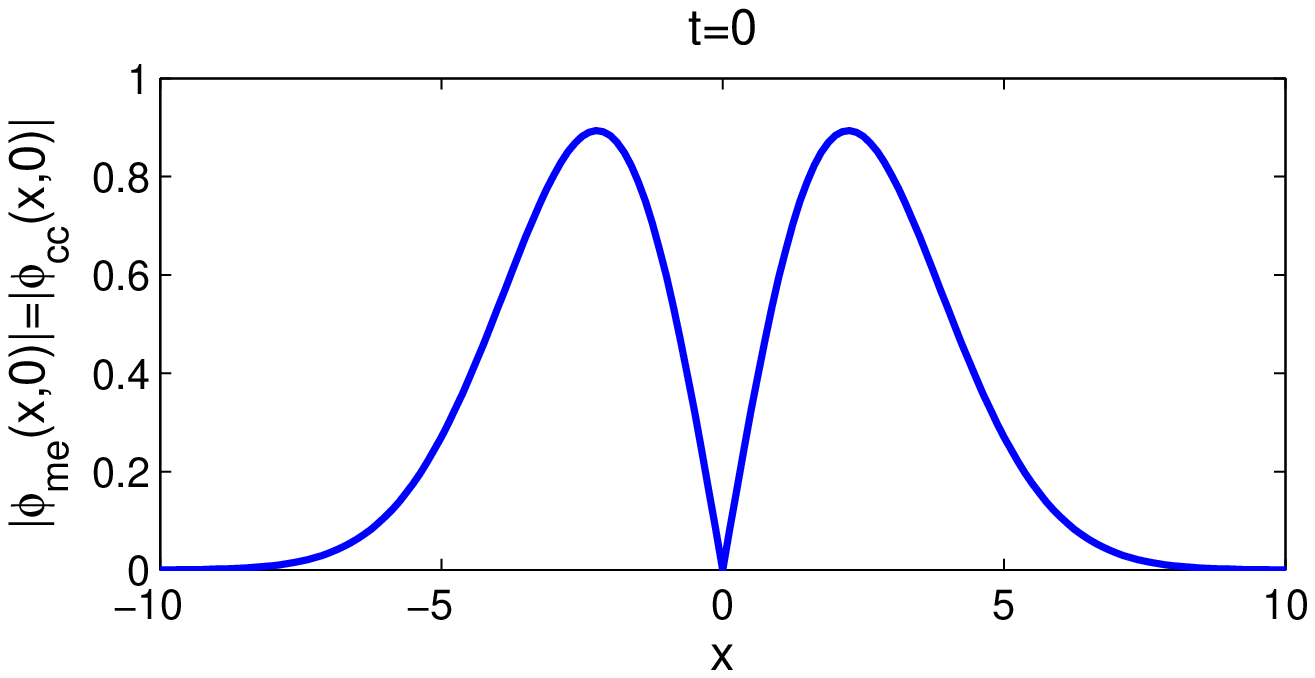,height=4.8cm,width=6.3cm}&\psfig{figure=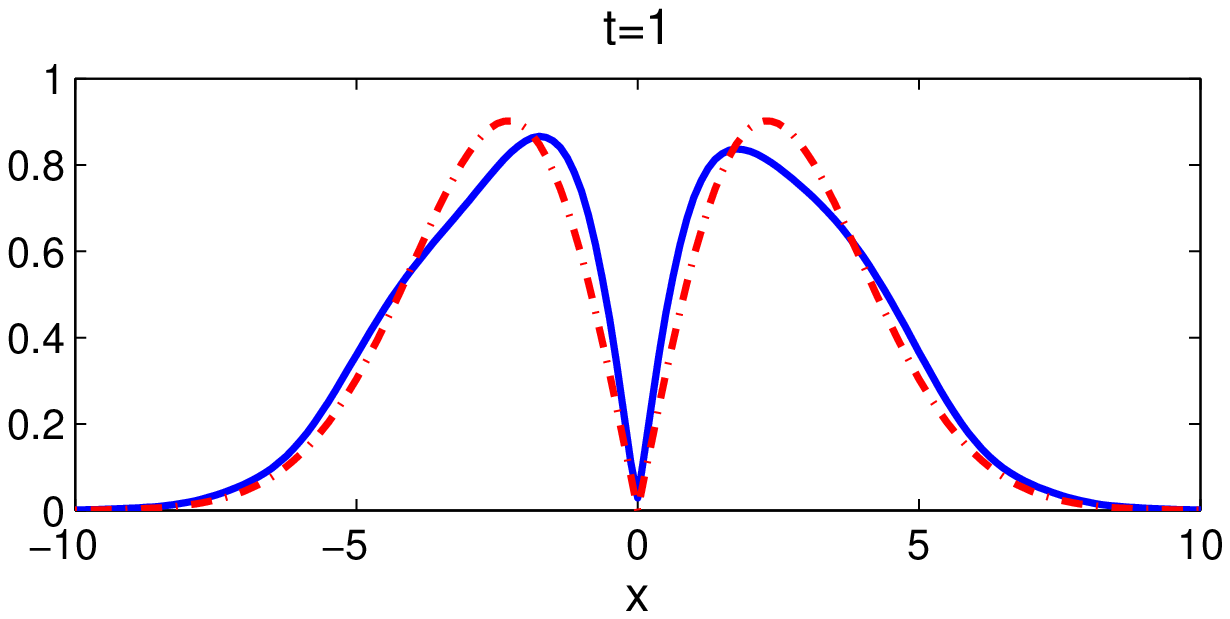,height=4.8cm,width=6.3cm}\\
\psfig{figure=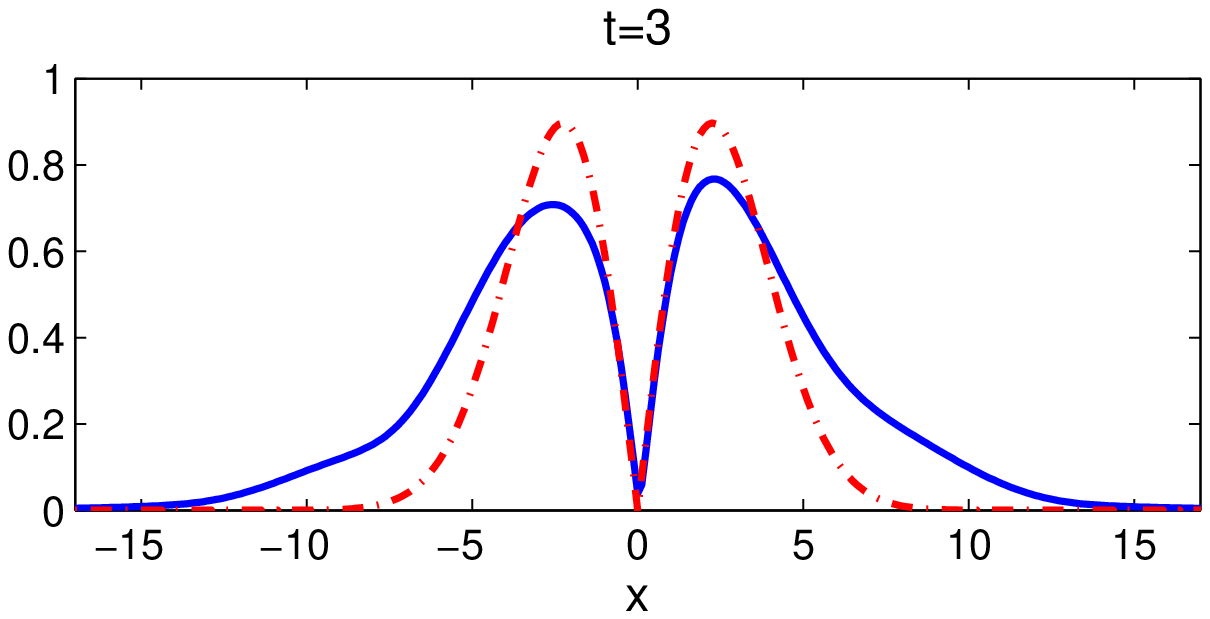,height=4.8cm,width=6.3cm}&\psfig{figure=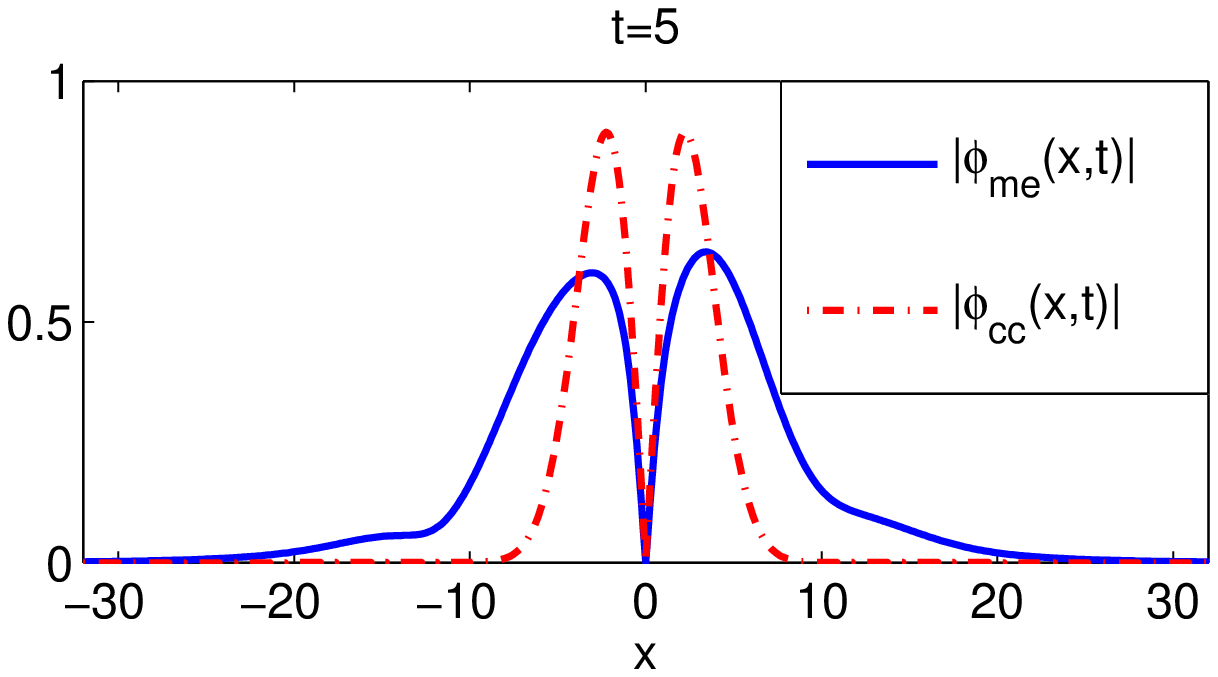,height=4.8cm,width=6.3cm}\\
\psfig{figure=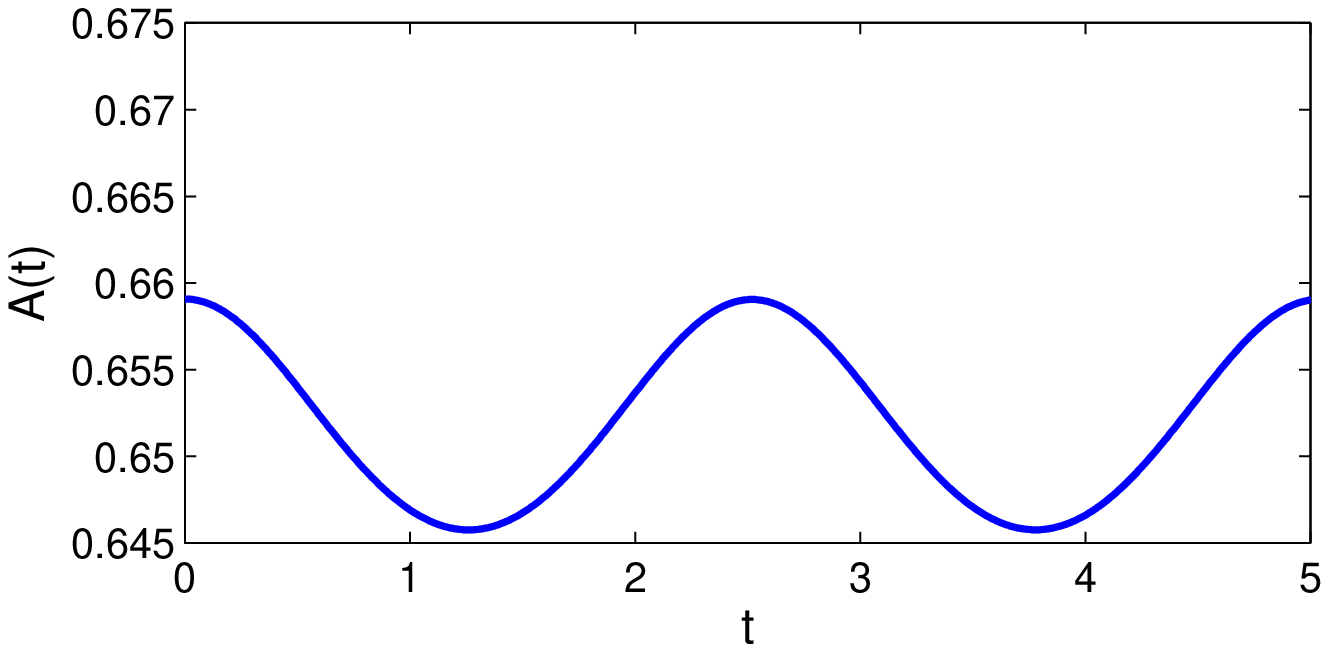,height=4.8cm,width=6.3cm}&\psfig{figure=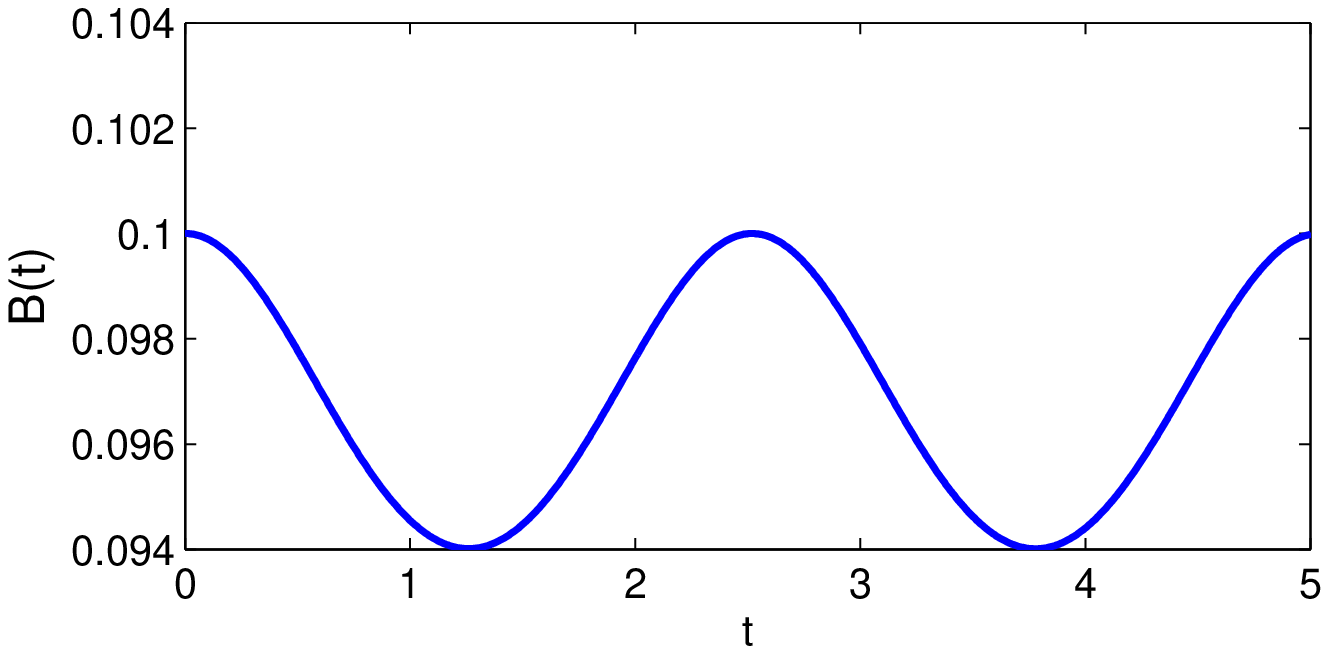,height=4.8cm,width=6.3cm}
\end{array}
$$
\caption{The radiation $|\phi_{me}(x,t)|$ given by modulation equation method and the $|\phi_{cc}(x,t)|=A(t)\fe^{-B(t)x^2}$ given by collective coordinates method at different $t$ in the comparison example (\ref{model}). The last row shows the dynamics of the amplitude $A(t)$ and width $B(t)$ from the collective coordinates method.}\label{figadd2}
\end{figure}

In the last but not least comparison, we apply the collective coordinates method to study the multichannel solution via solving the model NLS problem
\begin{subequations}\label{model}
\begin{eqnarray}
&i\partial_t\Phi(x,t)=\left[-\partial_{xx}+V(x)+\lambda|\Phi|^{2}\right]\Phi(x,t),\quad x\in\bR,\ t>0,\\
&\Phi(x,0)=\fe^{i\gamma_0}\psi_{E_0}(x)+\phi_0(x),\quad x\in\bR,
\end{eqnarray}
\end{subequations}
where the potential, initial data, parameters are chosen same as (\ref{num: ini0}) and (\ref{num: ini}), and the initial radiation
$$\phi_0(x)=\fe^{i\alpha_0-B_0x}A_0x,\quad \mbox{with}\quad\alpha_0=-0.5,\ B_0=0.1,\ A_0=0.66.$$
The above model NLS is associated with the Lagrangian density
$$\mathcal{L}:=\frac{i}{2}(\Psi\overline{\Psi}_t-\overline{\Psi}\Psi_t)+|\Psi_x|^2+V(x)|\Psi|^2+\frac{\lambda}{2}|\Psi|^4,$$
and the averaged Lagrangian $L:=\int_\bR\mathcal{L}dx$. As usually used in the literatures \cite{Alawi,Alamoudi,Collective2,BBB,Dawes,Morales1,374} and in particular \cite{Malomed1,Malomed} where special attention has been paid to the interaction of solitons and radiation, the collective coordinates method takes the ansatz
\begin{equation}
\Psi(x,t)=\fe^{-i\theta(t)}\psi_{E(t)}(x)+\fe^{i\alpha(t)}A(t)x\fe^{-B(t)x^2},\quad x\in\bR,\ t>0,\label{cc ansatz}
\end{equation}
with $\psi_{E(t)}$ the nonlinear bound state and five real-valued time dependent parameters $\theta(t)$, $E(t)$, $\alpha(t)$, $B(t)$ and $A(t)$. Plugging (\ref{cc ansatz}) into the Lagrangian $L$ and 
 noting
$$E\int\psi_E^2dx=\int[(\partial_x\psi_E)^2+V\psi_E^2+\lambda\psi_E^4]dx,$$
we get
\begin{eqnarray}
L=&(-\dot{\theta}+E)\int_\bR\psi_{E}^2dx-\frac{\lambda}{2}\int_\bR\psi_{E}^4dx+A^2\int_\bR(V+\dot{\alpha})x^2\fe^{-2Bx^2}dx\nonumber\\
&+A^2\int_\bR(1-2Bx^2)^2\fe^{-2Bx^2}dx+\frac{\lambda}{2} A^4\int_\bR x^4\fe^{-4Bx^2}dx\nonumber\\
&+\lambda A^2[1+2\cos^2(\theta+\alpha)]\int_\bR x^2\psi_{E}^2\fe^{-2Bx^2}dx.
\end{eqnarray}
Then by the Euler-Lagrangian equations from the variational principle, we get
\begin{subequations}
\begin{eqnarray}\label{reduced law}
&\dot{\theta}=E+\lambda A^2[1+2\cos^2(\theta+\alpha)]\frac{\int_\bR x^2\psi_E\partial_E\psi_E\fe^{-2Bx^2}dx}{\int_\bR \psi_E\partial_E\psi_Edx},\quad t>0,\\
& \dot{E}=\lambda A^2\sin(2(\theta+\alpha))\frac{\int_\bR x^2\psi_E^2\fe^{-2Bx^2}dx}{\int_\bR \psi_E\partial_E\psi_Edx},\\
&\dot{\alpha}=-\frac{3\lambda A^2}{4B}-3B-\frac{4\sqrt{2}}{\sqrt{\pi}}\lambda B^{\frac{3}{2}}[1+2\cos^2(\theta+\alpha)]\int_\bR x^2\psi_E^2\fe^{-2Bx^2}dx\nonumber\\
&\quad\ \ \,-\frac{4\sqrt{2}}{\sqrt{\pi}}B^{\frac{3}{2}}\int_\bR Vx^2\fe^{-2Bx^2}dx,\\
&\frac{\sqrt{\pi}}{4\sqrt{2}}B^{-\frac{3}{2}}\dot{A}-\frac{3\sqrt{\pi}}{16\sqrt{2}}B^{-\frac{5}{2}}A\dot{B}
=-\lambda A\sin(2(\theta+\alpha))\int_\bR x^2\psi_E^2\fe^{-2Bx^2}dx,\\
&\lambda c_0A^2=\frac{3\sqrt{\pi}}{4\sqrt{2}}B^2+B^{\frac{7}{2}}
\int_\bR V\fe^{-2Bx^2}\left(x^4-\frac{3}{4 B}x^2\right)dx\nonumber\\
&\qquad\quad\ +\lambda B^{\frac{7}{2}}[1+2\cos^2(\theta+\alpha)]\int_\bR \psi_E^2\fe^{-2Bx^2}\left(x^4-\frac{3}{4 B}x^2\right)dx,\label{constraint}
\end{eqnarray}
\end{subequations}
where $c_0=\left(\frac{9}{64\sqrt{2}}-\frac{15}{1024}\right)\sqrt{\pi}$, with initial data
$$A(0)=A_0,\quad B(0)=B_0,\quad \alpha(0)=\alpha_0,\quad E(0)=E_0,\quad \theta(0)=-\gamma_0.$$
Solving the above differential equations, we get the dynamics of the parameters predicted by the collective coordinates method. Fig. \ref{figadd1} shows the results related to the soliton, i.e. $E$ and $\theta$, together with comparisons with the results given by the modulation equation method, and Fig. \ref{figadd2} shows the radiation part.

From Figs. \ref{figadd1}\&\ref{figadd2}, we can see that: the predications given by the collective coordinates methods by using ansatz (\ref{cc ansatz}) are close to the exact solution in short time, especially in the soliton part. As time grows, the error from the predication tends to increases. This is mainly caused by the poor predication in the radiation at large time. As one can see in the last row of Fig. \ref{figadd2}, the predicated radiation turns to move periodically which is not correct. The error from the radiation part will finally break the soliton and stop us from reaching the steady state.
In fact unlike solitons, it is hard to capture the pattern of the radiation by finitely many variables in the collective coordinates method. In contrast, the modulation equation approach treats the radiation term exactly.

\section*{Acknowledgements}
A. Soffer is partially supported by NSF grant DMS– 1201394.
This work was done when the second author was visiting Department of Mathematics, Rutgers University, New Jersey, 2013. The authors would like to thank the referees for the helpful suggestions that greatly improved the paper.


\end{document}